 \documentclass[final,3p,times,twocolumn]{elsarticle}

\usepackage{amssymb}

\biboptions{sort&compress}

\journal{NIM B}

\begin{document}

\begin{frontmatter}



\title{Activation cross-sections of deuteron induced reactions on $^{nat}$Sm up to 50 MeV }


\author[1]{F. T\'ark\'anyi}
\author[2]{A. Hermanne}
\author[1]{S. Tak\'acs}
\author[1]{F. Ditr\'oi\corref{*}}
\author[1]{J. Csikai} 
\author[4]{A.V. Ignatyuk}
\cortext[*]{Corresponding author: ditroi@atomki.hu}

\address[1]{Institute for Nuclear Research, Hungarian Academy of Sciences (ATOMKI),  Debrecen, Hungary}
\address[2]{Cyclotron Laboratory, Vrije Universiteit Brussel (VUB), Brussels, Belgium}
\address[4]{Institute of Physics and Power Engineering (IPPE), Obninsk, Russia}

\begin{abstract}
Activation cross-sections for deuteron induced reactions on Sm  are presented for the first time  for $^{nat}$Sm(d,xn)$^{155,154,152m2,152m1,152g,150m,150g,149,148,147,146}$Eu, $^{nat}$Sm(d,x)$^{153,145}$Sm and $^{nat}$Sm(d,x)$^{151,150,149,145,144,143}$Pm up to 50 MeV. The cross-sections were measured by the stacked-foil irradiation technique and high resolution $\gamma$-ray spectrometry. The results were compared with results of nuclear reaction codes ALICE-D, EMPIRE-D and TALYS (from TENDL libraries). Integral yields of the products were calculated from the excitation functions.
\end{abstract}


\end{frontmatter}


\section{Introduction}
\label{1}
In the frame of a systematic study of activation cross-sections for deuteron induced reactions, we have investigated the excitation functions for $\gamma$-emitting radionuclides produced on natural samarium up to 50 MeV.
The aim of the program was getting reliable experimental data for different applications (medical isotope production, thin layer activation technique for wear measurement, radiation safety, targetry, etc.) and for improvement the presently used nuclear reaction models. More details on the program and the investigated nuclear reactions can be found in our recent summary papers and conference reports \cite{1,2}.
In a thorough literature search no earlier data were found on production cross-sections of deuteron induced nuclear reactions on samarium. Only a few thick target yields at 22 MeV are available from Dmitriev \cite{3}.
Out of the presently measured data, activation cross-sections for the two medically relevant  reaction products $^{145}$Sm and $^{153}$Sm were published separately with a comparison of different production routes \cite{4}. For completeness we present here the measured data for these two radionuclides again in graphical form without discussion.

\section{Experiment and data evaluation}
\label{2}
The activation cross-sections were measured by using the well-known stacked foil activation method followed by high resolution $\gamma$-spectrometry of the irradiated target and beam monitor foils. The beam parameters along the stack were controlled by simultaneous re-measurement of the monitor reactions along the whole stack. So called elemental cross-sections were deduced considering the natural Sm as monoisotopic. The parameters and methods of experiment and data evaluation are summarized in Table 1. The used decay data and the contributing reactions with their Q-values are collected in Table 2. The cross-sections of the simultaneously measured monitor reactions ($^{27}$Al(d,x)$^{22,24}$Na  and $^{nat}$Ti(d,x)$^{48}$V ) reproduced  completely, both in shape and magnitude, the recommended excitation functions are shown in our earlier publication \cite{4}. In Table 2 the main $\gamma$-lines used for the evaluation are indicated in bold. These are mainly independent lines, but also the other listed $\gamma$-lines were also used if necessary.

\begin{table*}[t]
\tiny
\caption{Main experimental parameters and methods of data evaluation}
\centering
\begin{center}
\begin{tabular}{|p{1.1in}|p{1.3in}|p{0.9in}|p{1.2in}|} \hline 
\textbf{Reaction} & \textbf{${}^{nat}$Sm(d,x)} & \textbf{Data evaluation} & \textbf{Method} \\ \hline 
 &  &  &  \\ \hline 
Incident particle & Deuteron  & Gamma spectra evaluation & Genie 2000 \cite{5}, Forgamma \cite{6} \\ \hline 
Method  & Stacked foil & Determination of beam intensity & Faraday cup (preliminary)\newline Fitted monitor reaction (final) \cite{7} \\ \hline 
Stack composition & Ho-Sm-Al-Ti, repeated 27 times & Decay data & NUDAT 2.6 \cite{8} \\ \hline 
Target and thickness  & ${}^{nat}$Sm foils, 25.14, 23.13,\newline 24.20 $\mu$m & Reaction Q-values & Q-value calculator \cite{9} \\ \hline 
Number of target foils\newline Isotopic abundance & 27\newline 144-3.1 \%-stable\newline 147-15.0 \%-1.06*10${}^{11}$ a 148-11.3 \%-7*10${}^{15}$ a \newline 149-13.8 \%-7*10${}^{15}$ a\newline 150-7.4 \%- stable\newline 152-26.7- stable\newline 154-22.7- stable & Determination of  beam energy & Andersen (preliminary) \cite{10}\newline Fitted monitor reaction (final) \cite{7}  \\ \hline 
Accelerator & Cyclone 90 cyclotron of the Université Catholique in CityplaceLouvain la Neuve (LLN)  & Uncertainty of energy & cumulative effects of possible uncertainties \\ \hline 
Primary energy & 50 MeV & Cross-sections & Elemental cross-section \\ \hline 
Irradiation time & 66 min & Uncertainty of cross-sections & sum in quadrature of all individual contributions \cite{11} \\ \hline 
Beam current & 120 nA & Yield & Physical yield \cite{12} \\ \hline 
Monitor reaction, [recommended values]  & ${}^{27}$Al(d,x)${}^{22,}$${}^{24}$Na \newline ${}^{nat}$Ti(d,x)${}^{48}$V reactions \cite{13} &  &  \\ \hline 
Monitor foils and thickness & ${}^{nat}$Al : 98 $\mu$m and 49.6 $\mu$m\newline ${}^{nat}$Ti ;  10.9 $\mu$m &  &  \\ \hline 
detector & HPGe &  &  \\ \hline 
g-spectra measurements & 4 series &  &  \\ \hline 
Cooling times & 6h, 25h, 70h, 530 h &  &  \\ \hline 
\end{tabular}
\end{center}

\end{table*}

\begin{table*}[t]
\tiny
\caption{Decay characteristics of the investigated activation products and Q-values of contributing reactions (bold gamma energies are the principal lines used for the evaluation)}
\centering
\begin{center}
\begin{tabular}{|p{0.8in}|p{0.6in}|p{0.6in}|p{0.6in}|p{0.9in}|p{0.7in}|} \hline 
Nuclide\newline Decay path & Half-life & E$_{\gamma}$(keV) & I$_{\gamma}$(\%) & Contributing reaction & Q-value\newline (keV) \\ \hline 
\textbf{${}^{155}$Eu\newline }~$\beta $${}^{-}$: 100~\%\textbf{} & 4.753 a~ & ~86.5479\newline 105.3083 & 30.7 \newline 21.1  & ${}^{154}$Sm(d,n) & 4427.03 \\ \hline 
\textbf{${}^{154g}$Eu\newline }~$\beta $${}^{-}$: 99.982\textbf{} & 8.601 a & ~123.0706\newline ~723.3014\newline 873.1834~\newline 996.29\newline 1004.76~\newline 1274.429 & 40.4 \newline 20.06 \newline 12.08 \newline 10.48 \newline 18.01\newline 34.8  & ${}^{154}$Sm(d,2n) & ~-3724.26 \\ \hline 
\textbf{${}^{152m2}$Eu\newline }~IT~\newline 147.81 keV\textbf{} & 96 min~ & ~89.847 & ~69.9  & ${}^{152}$Sm(d,2n)\newline ${}^{154}$Sm(d,4n) & ~-4881.52${}^{*}$\newline -18716.71${}^{*}$ \\ \hline 
\textbf{${}^{152m1}$Eu\newline }$\beta $${}^{-}$: 72\textit{~}\%\newline $\varepsilon $: 28\textit{~}\%~\newline 45.5998\textit{ keV}\textbf{} & 9.3116 h & 344.31\newline 121.777\newline 841.594\newline 963.390 & 2.4 \newline ~7.0 ~\newline 14.2 \newline 11.7 ~ & ${}^{152}$Sm(d,2n)\newline ${}^{154}$Sm(d,4n) & ~-4881.52${}^{*}$\newline -18716.71${}^{*}$ \\ \hline 
\textbf{${}^{152g}$Eu\newline }$\beta $${}^{-}$: 27.9\textit{ }\%~\newline ~$\varepsilon $: 72.1\textit{ }\%~\textbf{} & 13.537 a & 344.2785\newline ~778.9040\newline 121.7817\newline 244.6975\newline 964.079\newline 1085.869\newline 1112.069\newline 1408.006 & ~26.6 \newline 12.96 \newline 28.67 \newline ~7.61 \newline 14.65 \newline 10.24 \newline 13.69 \newline ~21.07  & ${}^{152}$Sm(d,2n)\newline ${}^{154}$Sm(d,4n) & ~-4881.52\newline -18716.71 \\ \hline 
\textbf{${}^{150m}$Eu\newline }$\varepsilon $: 11 \%\newline IT 88.8 \%\newline 42.1~keV\textbf{} & 12.8 h & 333.9\newline ~\textbf{406.5} & ~4.0 \newline ~2.8  & ${}^{149}$Sm(d,n)\newline ${}^{15}$${}^{0}$Sm(d,2n)\newline ${}^{152}$Sm(d,4n)\newline ${}^{154}$Sm(d,6n) & ~2721.13${}^{*}$\newline ~-5265.55${}^{*}$\newline -18716.71${}^{*}$\newline ~-32954.9${}^{*}$ \\ \hline 
\textbf{${}^{150g}$Eu\newline }$\varepsilon $: 100 \%~\textbf{} & 36.9 a & \textbf{333.971\newline }~439.401\newline ~584.274\newline 1049.043 & ~96 \newline 80 \newline ~52.6 \newline ~5.4  & ${}^{149}$Sm(d,n)\newline ${}^{15}$${}^{0}$Sm(d,2n)\newline ${}^{152}$Sm(d,4n)\newline ${}^{154}$Sm(d,6n) & ~2721.13\newline ~-5265.55\newline ~-19119.7\newline ~-32954.9 \\ \hline 
\textbf{${}^{149}$Eu\newline }$\varepsilon $: 100 \%~ & 93.1 d & ~277.089\newline \textbf{327.526\underbar{}} & 3.56 \newline 4.03  & ${}^{148}$Sm(d,n)\newline ${}^{149}$Sm(d,2n)\newline ${}^{15}$${}^{0}$Sm(d,3n)\newline ${}^{152}$Sm(d,5n)\newline ${}^{154}$Sm(d,7n) & 2168.82\newline ~-3701.52\newline -11688.21\newline ~-25542.36\newline -39377.55 \\ \hline 
\textbf{${}^{148}$Eu\newline }$\varepsilon $: 100 \%~\textbf{} & 54.5 d & 414.028\newline 414.057\newline \textbf{550.284\newline }~553.231\newline 553.260\newline 571.962\newline 611.293\newline 629.987\newline 725.673 & 10.3 \newline ~10.1 \newline ~99 \newline 12.9 \newline 5.0 \newline ~9.6 \newline ~20.5 \newline 71.9 \newline 12.7  & ${}^{147}$Sm(d,n)\newline ${}^{148}$Sm(d,2n)\newline ${}^{149}$Sm(d,3n)\newline ${}^{15}$${}^{0}$Sm(d,4n)\newline ${}^{152}$Sm(d,6n)\newline ${}^{154}$Sm(d,8n) & ~2097.8\newline ~-6043.6\newline ~-11914.\newline -19900.6\newline ~-33754.8\newline -47590.0 \\ \hline 
\textbf{${}^{147}$Eu\newline }$\varepsilon $: 99.9978 \% \textit{ }\textbf{} & 24.1 d & ~121.220\newline 197.299\newline \textbf{677.516\newline }1077.043 & ~21.2 \newline 24.4 \newline 9.0 \newline ~5.69  & ${}^{147}$Sm(d,2n)\newline ${}^{148}$Sm(d,3n)\newline ${}^{149}$Sm(d,4n)\newline ${}^{15}$${}^{0}$Sm(d,5n)\newline ${}^{152}$Sm(d,7n)\newline ${}^{154}$Sm(d,9n) & ~-4728.53\newline -12869.91\newline -18740.25\newline -26726.94\newline -40581.09\newline -54416.27 \\ \hline 
\textbf{${}^{146}$Eu\newline }~$\varepsilon $: 100 \%~\textbf{} & 4.59 d & 633.083\newline ~634.137\newline \textbf{747.159} & ~35.9 \newline 45.0 \newline 99  & ${}^{147}$Sm(d,3n)\newline ${}^{148}$Sm(d,4n)\newline ${}^{149}$Sm(d,5n)\newline ${}^{15}$${}^{0}$Sm(d,6n)\newline ${}^{152}$Sm(d,8n)\newline ${}^{154}$Sm(d,10n) & -13226.83\newline -21368.2\newline ~-27238.55\newline -35225.23\newline ~-49079.38\newline  \\ \hline 
\textbf{${}^{145}$Eu\newline }~$\varepsilon $: 100 \%~\textbf{} & 5.93 d & ~653.512\newline \textbf{~893.73\newline }~1658.53 & 15.0 \newline 66 \newline 14.9  & ${}^{144}$Sm(d,n)\newline ${}^{147}$Sm(d,4n)\newline ${}^{148}$Sm(d,5n)\newline ${}^{149}$Sm(d,6n)\newline ${}^{15}$${}^{0}$Sm(d,7n)\newline ${}^{152}$Sm(d,9n) & 1090.5~\newline -20424.02\newline ~-28565.4\newline ~-34435.7~\newline ~-42422.43\newline ~-56276.58 \\ \hline 
\textbf{${}^{153}$Sm\newline }$\beta $${}^{-}$: 100 \%\textbf{} & 46.50 h & \textbf{103.18012\newline } & 29.25  & ${}^{152}$Sm(d,p)\newline ${}^{154}$Sm(d,p2n)\newline ${}^{153}$Pm decay  & 3643.834\newline ~~-10191.4\newline -11320.47 \\ \hline 
\textbf{${}^{145}$Sm\newline }$\varepsilon $: 100 \%~\textbf{} & 340 d & \textbf{61.2265} & 12.15  & ${}^{144}$Sm(d,p)\newline ${}^{147}$Sm(d,p3n)\newline ${}^{148}$Sm(d,p4n)\newline ${}^{149}$Sm(d,p5n)\newline ${}^{15}$${}^{0}$Sm(d,p6n)\newline ${}^{152}$Sm(d,p8n)\newline ${}^{154}$Sm(d,p10n)\newline ${}^{145}$Eu decay\newline ${}^{145}$Pm decay & 4532.534\newline -16981.99\newline -25123.37\newline -30993.72\newline -38980.41\newline -52834.55~\newline -45532.\newline ~-15583.55 \\ \hline 
\textbf{${}^{151}$Pm\newline }~$\beta $${}^{-}$: 100 \%~\textbf{} & 28.40 h & 240.09\newline \textbf{340.08\newline }445.68 & 3.8 \newline 22.5  \newline 4.0 \newline  & ${}^{152}$Sm(d,2pn)\newline ${}^{154}$Sm(d,2p3n)\newline  & -10890.12\newline ~-24725.31 \\ \hline 
\textbf{${}^{150}$Pm\newline }$\beta $${}^{-}$:100 \%~\textbf{} & 2.68 h & \textbf{333.92\newline }~406.51\newline 831.85\newline 876.41\newline 1165.77\newline 1324.51~ & ~68 \newline ~5.6 \newline ~11.9 \newline 7.3 \newline 15.8\newline 17.5 & ${}^{15}$${}^{0}$Sm(d,2p)\newline ${}^{152}$Sm(d,2p2n)\newline ${}^{154}$Sm(d,2p4n)\newline  & ~~-4896.2\newline ~-18750.4\newline -32585.6 \\ \hline 

\end{tabular}
\end{center}
\end{table*}

\setcounter{table}{1}
\begin{table*}[t]
\tiny
\caption{Table 2. (continued) }
\centering
\begin{center}
\begin{tabular}{|p{0.8in}|p{0.6in}|p{0.6in}|p{0.6in}|p{0.9in}|p{0.7in}|} \hline 
Nuclide\newline Decay path & Half-life & E$_{\gamma}$(keV) & I$_{\gamma}$(\%) & Contributing reaction & Q-value\newline (keV) \\ \hline 
\textbf{$^{149}$Pm\newline }$\beta $$^{-}$: 100~\%~ & 53.08~h & 285.95 & 3.1~ & ${}^{149}$Sm(d,2p)\newline ${}^{150}$Sm(d,2pn)\newline ${}^{152}$Sm(d,2p3n)\newline $^{154}$Sm(d,2p5n) & -2513.63\newline -10500.32\newline -24354.47\newline -38189.66 \\ \hline 
\textbf{${}^{155}$Eu\newline }~$\beta $${}^{-}$: 100~\%\textbf{} & 4.753 a~ & ~86.5479\newline 105.3083 & 30.7 \newline 21.1  & ${}^{154}$Sm(d,n) & 4427.03 \\ \hline 
\textbf{${}^{154g}$Eu\newline }~$\beta $${}^{-}$: 99.982\textbf{} & 8.601 a & ~123.0706\newline ~723.3014\newline 873.1834~\newline 996.29\newline 1004.76~\newline 1274.429 & 40.4 \newline 20.06 \newline 12.08 \newline 10.48 \newline 18.01\newline 34.8  & ${}^{154}$Sm(d,2n) & ~-3724.26 \\ \hline 
\textbf{${}^{152m2}$Eu\newline }~IT~\newline 147.81 keV\textbf{} & 96 min~ & ~89.847 & ~69.9  & ${}^{152}$Sm(d,2n)\newline ${}^{154}$Sm(d,4n) & ~-4881.52${}^{*}$\newline -18716.71${}^{*}$ \\ \hline 
\textbf{${}^{148m}$Pm\newline }$\beta $${}^{-}$: 100~\%\newline 137.9 keV\textbf{} & 41.29~d & ~288.11\newline 414.07\newline 432.745\newline ~550.284~\newline \textbf{629.987\newline }725.673\newline 915.331~~\newline 1013.808~ & 12.56 \newline 18.66 \newline 5.33~\newline 94.5~\newline 89~\newline 32.7\newline 17.10~\newline 20.20~ & ${}^{148}$Sm(d,2p)\newline ${}^{149}$Sm(d,2pn)\newline ${}^{150}$Sm(d,2p2n)\newline ${}^{152}$Sm(d,2p4n)\newline ${}^{154}$Sm(d,2p6n)\newline  & ~-3913.21${}^{*}$\newline -9783.56${}^{*}$\newline ~-17770.25${}^{*}$\newline ~~-31624.4${}^{*}$\newline ~-45459.59${}^{*}$ \\ \hline 
\textbf{${}^{148g}$Pm\newline }$\beta $${}^{-}$: 100 \%~\textbf{} & 5.368 d & ~550.27\newline ~914.85\newline 1465.12 & 22.0 \newline 11.5 \newline 22.2  & ${}^{148}$Sm(d,2p)\newline ${}^{149}$Sm(d,2pn)\newline ${}^{15}$${}^{0}$Sm(d,2p2n)\newline ${}^{152}$Sm(d,2p4n)\newline ${}^{154}$Sm(d,2p6n)\newline  & ~-3913.21\newline -9783.56\newline ~-17770.25\newline ~~-31624.4\newline ~-45459.59 \\ \hline 
\textbf{${}^{146}$Pm\newline } $\varepsilon $: 66.0\textit{ }\%~\textit{\newline }~$\beta $${}^{-}$: 34.0~\%\textbf{} & ~5.53 a & \textbf{453.88\newline }~735.93 & 65.0 \newline 22.5  & ${}^{147}$Sm(d,2pn)\newline ${}^{148}$Sm(d,2p2n)\newline ${}^{149}$Sm(d,2p3n)\newline ${}^{15}$${}^{0}$Sm(d,2p4n)\newline ${}^{152}$Sm(d,2p6n)\newline ${}^{154}$Sm(d,2p8n)\newline  & ~-9325.34\newline -17466.72\newline ~-23337.07\newline -31323.76\newline -37518.85\newline ~-59013.09 \\ \hline 
\textbf{${}^{1}$${}^{44}$Pm\newline }$\varepsilon $: 100 \%~\textbf{} & 363 d & ~~476.78\newline 618.01\newline \textbf{696.49} & ~43.8 \newline ~98 \newline 99.490 & ${}^{147}$Sm(d,2p3n)\newline ${}^{148}$Sm(d,2p4n)\newline ${}^{149}$Sm(d,2p5n)\newline ${}^{15}$${}^{0}$Sm(d,2p6n)\newline ${}^{152}$Sm(d,2p8n) & -23506\newline -31647\newline -37517\newline -45504\newline -59538 \\ \hline 
\textbf{${}^{1}$${}^{43}$Pm\newline }$\varepsilon $: 100 \%~\textbf{} & ~265 d & \textbf{~741.98} & ~38.5  & ${}^{147}$Sm(d,2p4n)\newline ${}^{148}$Sm(d,2p5n)\newline ${}^{149}$Sm(d,2p6n)\newline ${}^{15}$${}^{0}$Sm(d,2p7n) & -30033.03\newline -38174.41\newline -44044.76\newline -52031.45 \\ \hline 
\end{tabular}
\end{center}

\begin{flushleft}
\footnotesize{\noindent Increase Q-values if compound particles are emitted: np-d, +2.2 MeV; 2np-t, +8.48 MeV; n2p-${}^{3}$He, +7.72 MeV; 2n2p-a, +28.30 MeV 

\noindent *Decrease Q-values for isomeric states with level energy of the isomer }

\end{flushleft}

\end{table*}

\section{Model calculations}
\label{3}
 For theoretical estimation the updated ALICE-IPPE-D \cite{14} and EMPIRE-D \cite{15} codes were used. \cite{16,17}. Results of ALICE-IPPE-D for excited states were obtained by applying the isomeric ratios derived from the EMPIRE-D code to the total cross-sections calculated by ALICE-IPPE-D.
 For a comparison of the experimental the widely available theoretical data from the TENDL-2012 \cite{18} library (based on the modified TALYS code \cite{19}) are also presented.

\section{Results and discussion}
\label{4}

\subsection{Cross sections}
\label{4.1}

The cross-sections for radioisotopes of Eu and Pm produced in the bombardment of $^{nat}$Sm with deuterons are tabulated in Tables 3, 4 and 5. The numerical values for $^{145}$Eu, 145,$^{153}$Sm can be found in our previous publication [4].  The comparison with the theory is shown in Figs 1-22.  We present the TENDL-2012 values up to 60 MeV to illustrate the tendency of the excitation functions above the experimentally measured energy range. The data for production of $^{145}$Eu, $^{145}$Sm and $^{153}$Sm are presented only in graphical form as the numerical data are available in \cite{4}. The contributing reactions and their Q-values are listed in Table 2. We do not explicitly discuss the contribution of the very low half-life isomeric states decaying completely by isomeric transition to the ground state.	

\subsubsection{Production of radioisotopes of europium}
\label{4.1.1}
\vspace{0.4 cm}
 \textbf{4.1.1.1	Production cross-sections of $^{155}$Eu}\\

The long-lived $^{155}$Eu (GS, J$^\pi$=5/2$^+$, T$_{1/2}$ = 4.753 a, $\beta^-$: 100 \%) is produced directly via (d,xn) reactions and from the $\beta^-$-decay of the short-lived parent $^{155}$Sm (22.3 min). As it was mentioned, the typical cooling time for the first series of $\gamma$-spectra measurement was around 6 h and hence we could not measure the $^{155}$Sm contribution separately. The cross-section of $^{155}$Eu is cumulative after complete decay of the $^{155}$Sm.
The experimental data are compared with the theoretical results in the Fig. 1 (both the direct and cumulative theoretical cross-section are presented). The TENDL-2012 values underestimate by a factor of two the cumulative cross-sections due to the known underestimation of the (d,p) process in the TALYS code. The agreement with ALICE-D and EMPIRE-D is acceptable good. The curve calculated by systematics \cite{20} gives also an acceptable estimation.

\begin{figure}
\includegraphics[scale=0.3]{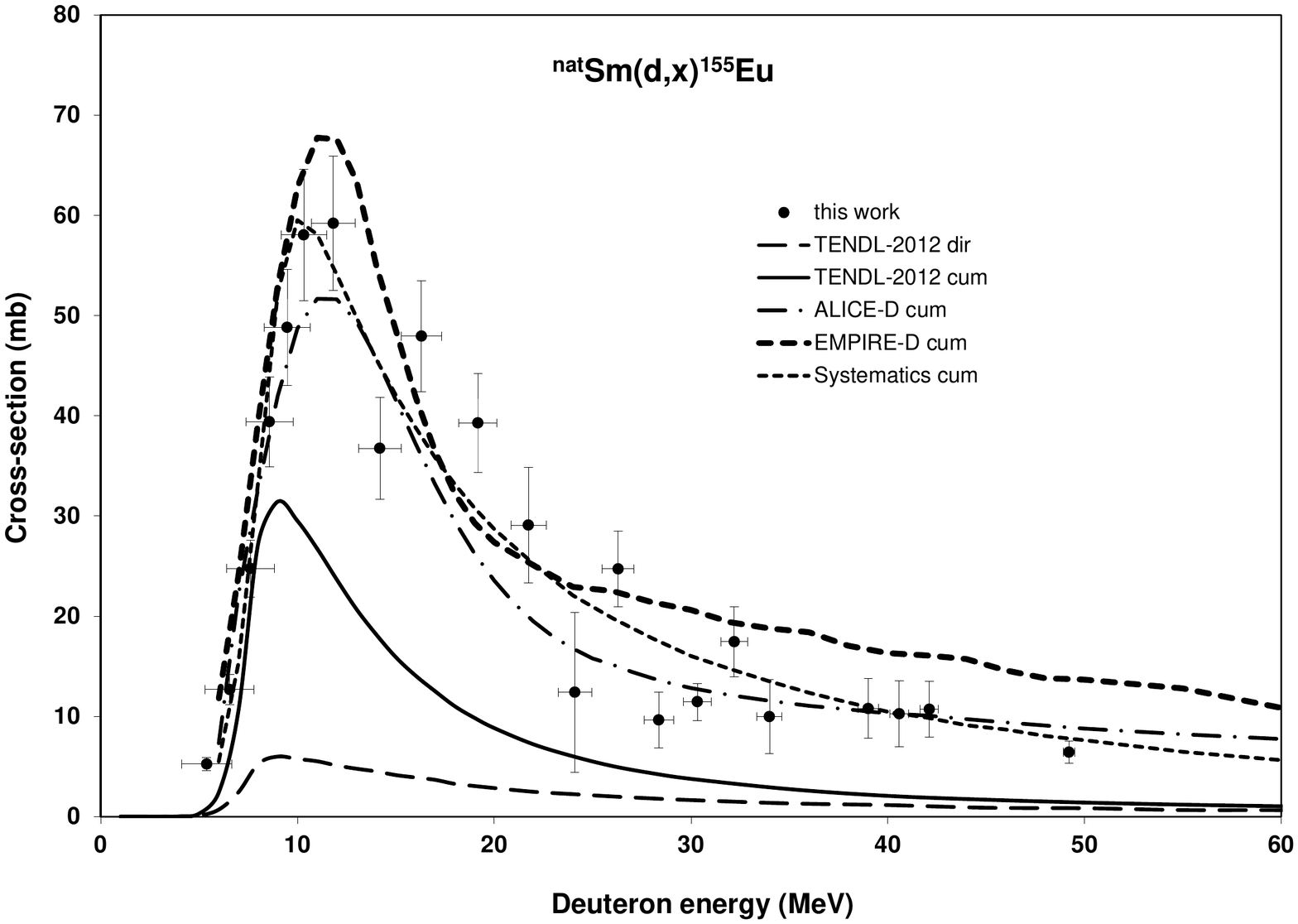}
\caption{Excitation functions of the $^{nat}$Sm(d,xn)$^{155}$Eu reaction in comparison with results from model calculations}
\end{figure}
\vspace{0.4 cm}
\textbf{4.1.1.2	Production cross-sections of $^{154}$Eu}\\

The production cross-sections of the $^{154}$Eu ground state (GS, J$^\pi$ = 3$^-$ , T$_{1/2}$ = 8.601 a, $\beta^-$: 99.982 \%, $\varepsilon$: 0.018 \%) contain the direct production and the contribution of complete decay of the short-lived  isomeric state (MS, E(level) = 145.3 keV,  J$^\pi$ = 8$^-$ , T$_{1/2}$ =   46.0 min, IT: 100 \%). The agreement between the experimental data and the TENDL-2012 and ALICE-D results (cumulative) for the production of $^{154}$Eu  (m+) is good below 15 MeV (Fig. 2). The overestimation of the EMPIRE-D is significant even in the lower energy region.

\begin{figure}
\includegraphics[scale=0.3]{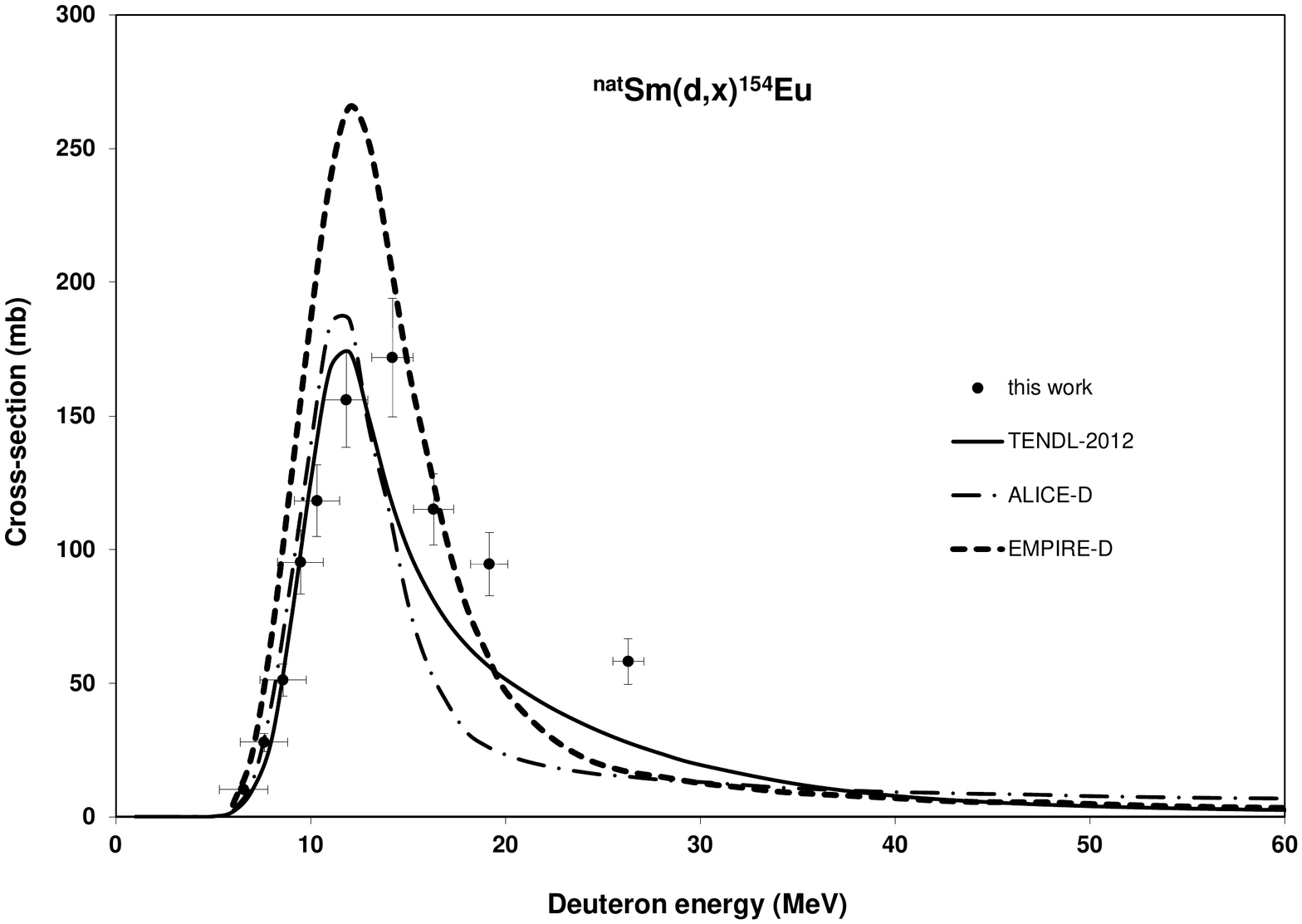}
\caption{Excitation functions of the $^{nat}$Sm(d,xn)$^{154}$Eu reaction in comparison with results from model calculations}
\end{figure}
\vspace{0.4 cm}
\textbf{4.1.1.3	Production cross-sections of different levels of  $^{152}$Eu }\\

The radionuclide $^{152}$Eu has apart from its long-lived ground state (GS, T$_{1/2}$ = 13.537 a, J$^\pi$ = 3$^-$, $\beta^-$: 27.9 \%, $\varepsilon$: 72.1 \%) two longer-lived isomeric states with complex decay scheme. The high spin, rather short-lived, isomer $^{152m2}$Eu (MS2, E(level) = 147.81 keV, J$^\pi$ =  8$^-$, T$_{1/2}$ = 96 min, IT  100 \%) decays totally by isomeric transition to the ground state, without population of the lower excited $^{152m1}$Eu level. This latter metastable state with zero spin (MS1, E(level) = 45.5998 keV, J$^\pi$ = 0$^-$, T$_{1/2}$ = 9.3116 h, $\beta^-$: 72 \%, $\varepsilon$: 28 \%) decays by $\beta$-decay to stable $^{152}$Gd and does not populate the ground state. Independent $\gamma$-rays are emitted in the decay of the different states so that production cross-section for all three can be determined  
\vspace{0.4 cm}
\textbf{4.1.1.3a	Production cross-sections of  $^{152m2}$Eu}\\

The results for the high spin isomer (MS2, E(level) = 147.81 keV, J$^\pi$ =  8$^-$, T$_{1/2}$ = 96 min, IT 100 \%) are shown in Fig. 3. Although in the experimental values the contribution of reactions on the two high mass stable target isotopes can clearly be seen, and the three codes represent well the shape and position of the maxima, there are large disagreements with the theoretical data, which are predicting much higher cross-sections. 

\begin{figure}
\includegraphics[scale=0.3]{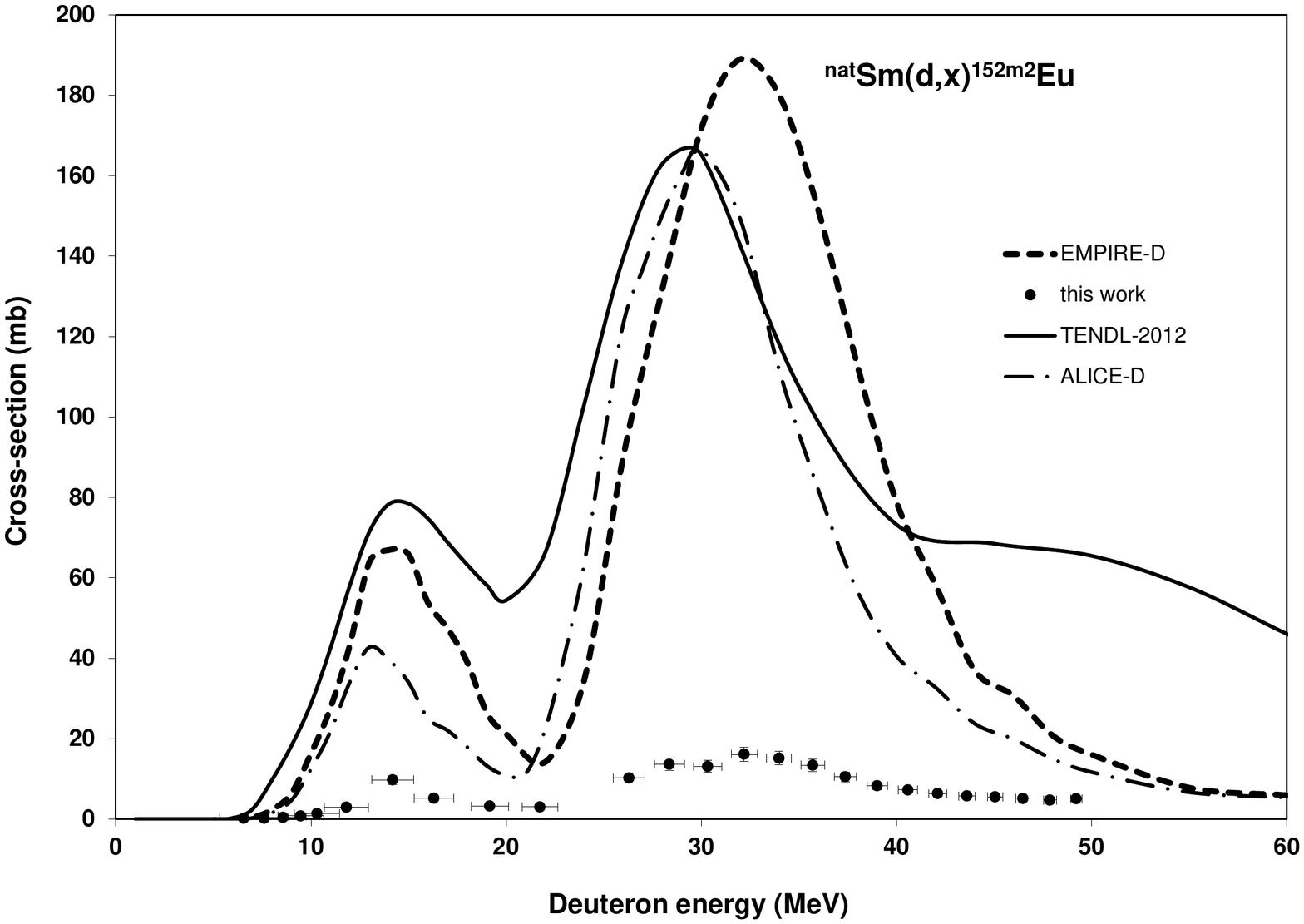}
\caption{Excitation functions of the $^{nat}$Sm(d,xn)$^{152m2}$Eu reaction in comparison with results from model calculations }
\end{figure}

\vspace{0.4 cm}
\textbf{4.1.1.3b	Production cross-sections of  $^{152m1}$Eu}\\

The results for the low spin isomer (MS1, E(level) = 45.5998 keV, J$^\pi$ = 0$^-$, T$_{1/2}$ =  9.3116 h, $\beta^-$: 72 \%, $\varepsilon$: 28 \%) are shown in Fig. 4. Experimentally the contribution of the reactions on $^{152}$Sm and $^{154}$Sm can be distinguished. The codes describe now rather well the experimental values, with TENDL-2012 values being about 50 \% too low. 

\begin{figure}
\includegraphics[scale=0.3]{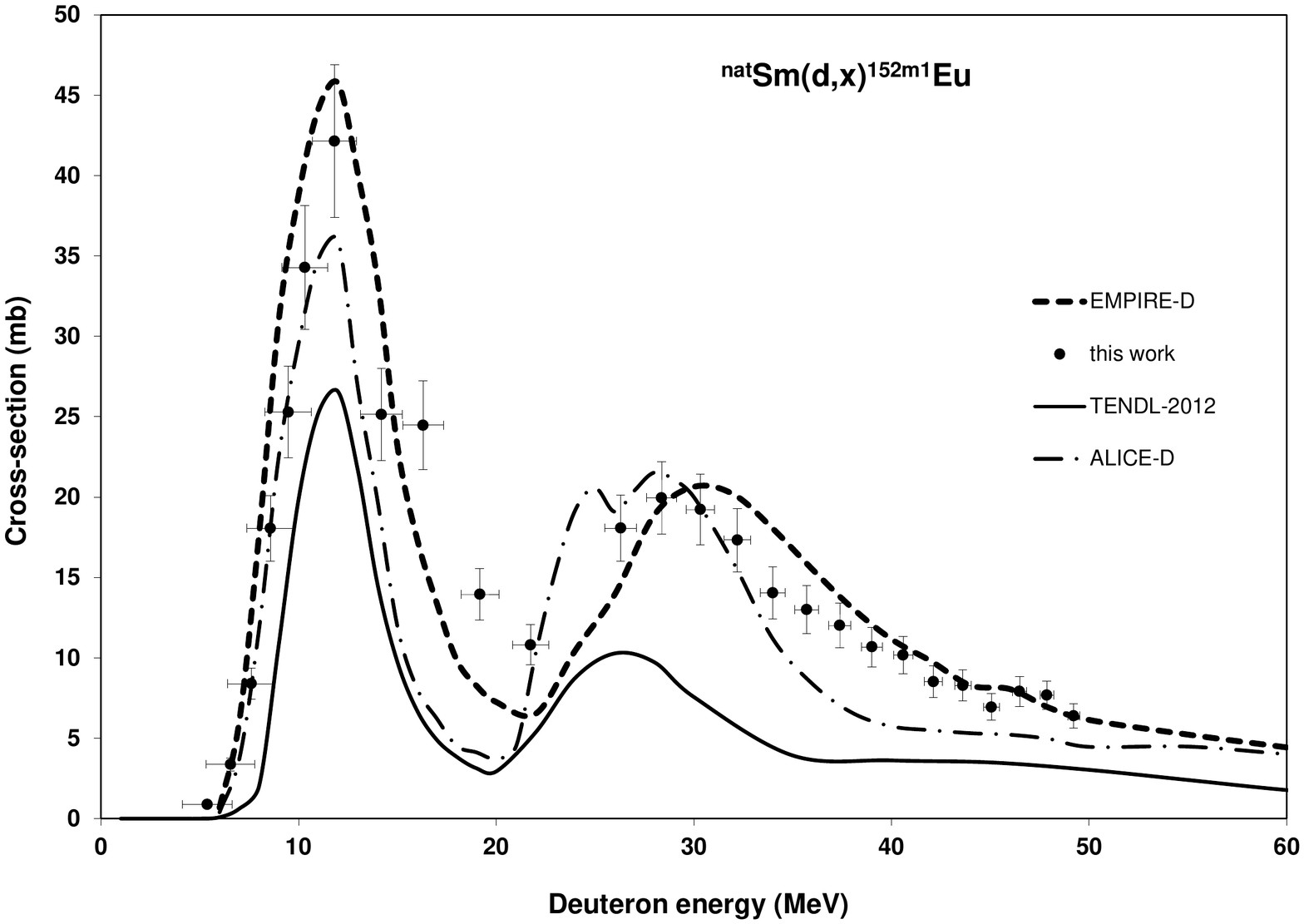}
\caption{Excitation functions of the $^{nat}$Sm(d,xn)$^{152m1}$Eu reaction in comparison with results from model calculations }
\end{figure}
\vspace{0.4 cm}
\textbf{4.1.1.3c	Production cross-sections of  $^{154g}$Eu}\\

The measured cross-sections for production of the ground state (GS, T$_{1/2}$ = 13.537 a, J$^\pi$ =3$^-$, $\beta^-$: 27.9 \%, $\varepsilon$: 72.1 \%) are shown in Fig. 5. As stated before, in the measurement taken at EOB + 70 h, and 530 h, in addition to direct production the contribution of total decay of  the high spin state $^{152m2}$Eu (IT: 100  \%) is also included (m2+).  In the energy region below 20 MeV ALICE-D and TENDL give acceptable estimation and EMPIRE-D overestimates around the first peak region. Above 25 MeV EMPIRE-D gives better result and the other two codes underestimate. 

\begin{figure}
\includegraphics[scale=0.3]{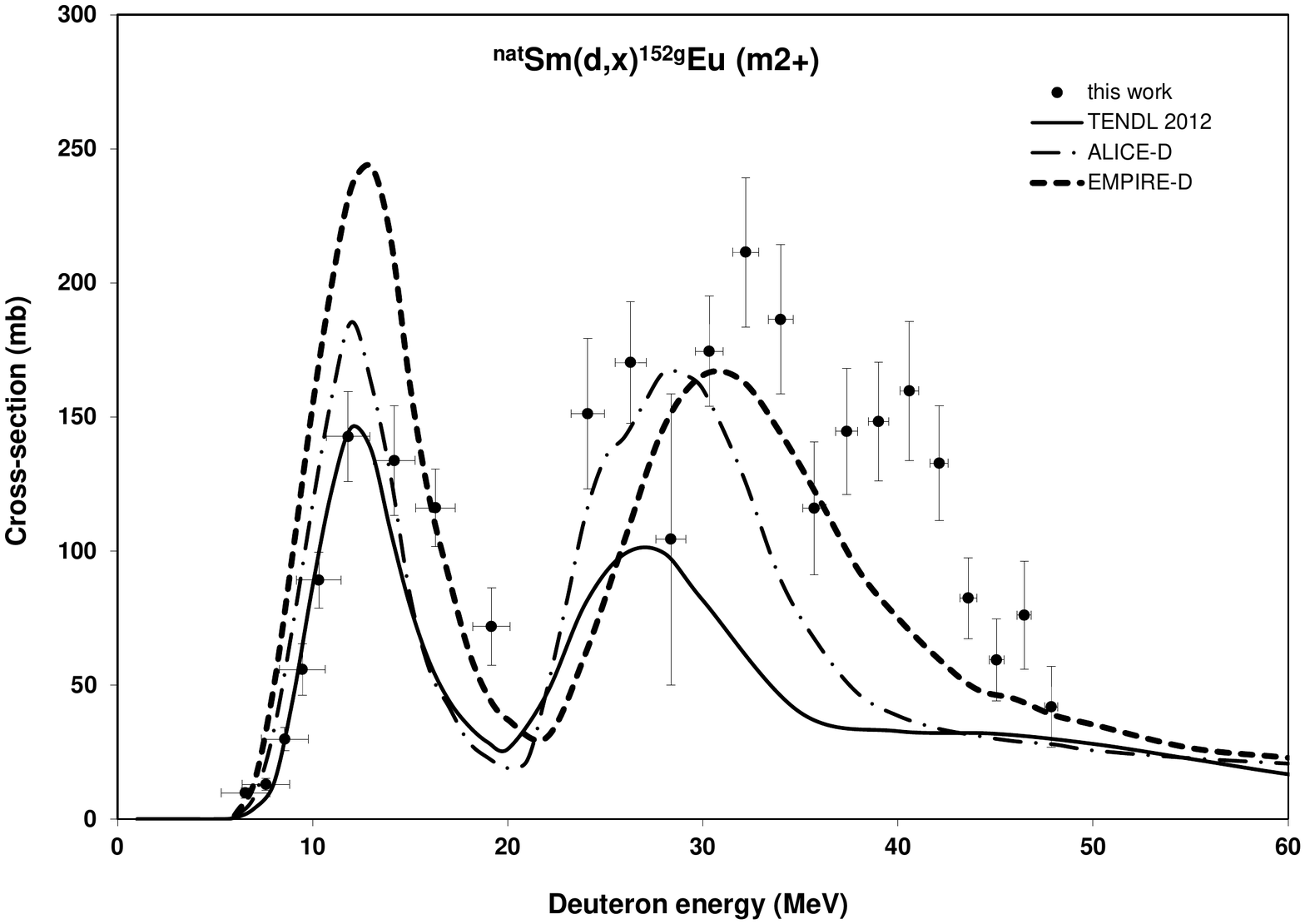}
\caption{Excitation functions of the $^{nat}$Sm(d,xn)$^{154g}$Eu (m2+) reaction in comparison with results from model calculations }
\end{figure}

\vspace{0.4 cm}
\textbf{4.1.1.4	Production cross-sections of  $^{150m}$Eu}\\

The cross-section of the long-lived isomeric state (MS, E(level) = 42.1 keV, J$^\pi$ =  0$^-$ T$_{1/2}$ = 12.8 h, $\varepsilon$: 11 \% , $\beta^-$: 89 \%, IT $<$ 5.*10$^{-8}$) is shown in Fig. 6. The three model codes reproduce the shape of the excitation function but overestimate the experimental data above 25 MeV deuteron energy.	

\begin{figure}
\includegraphics[scale=0.3]{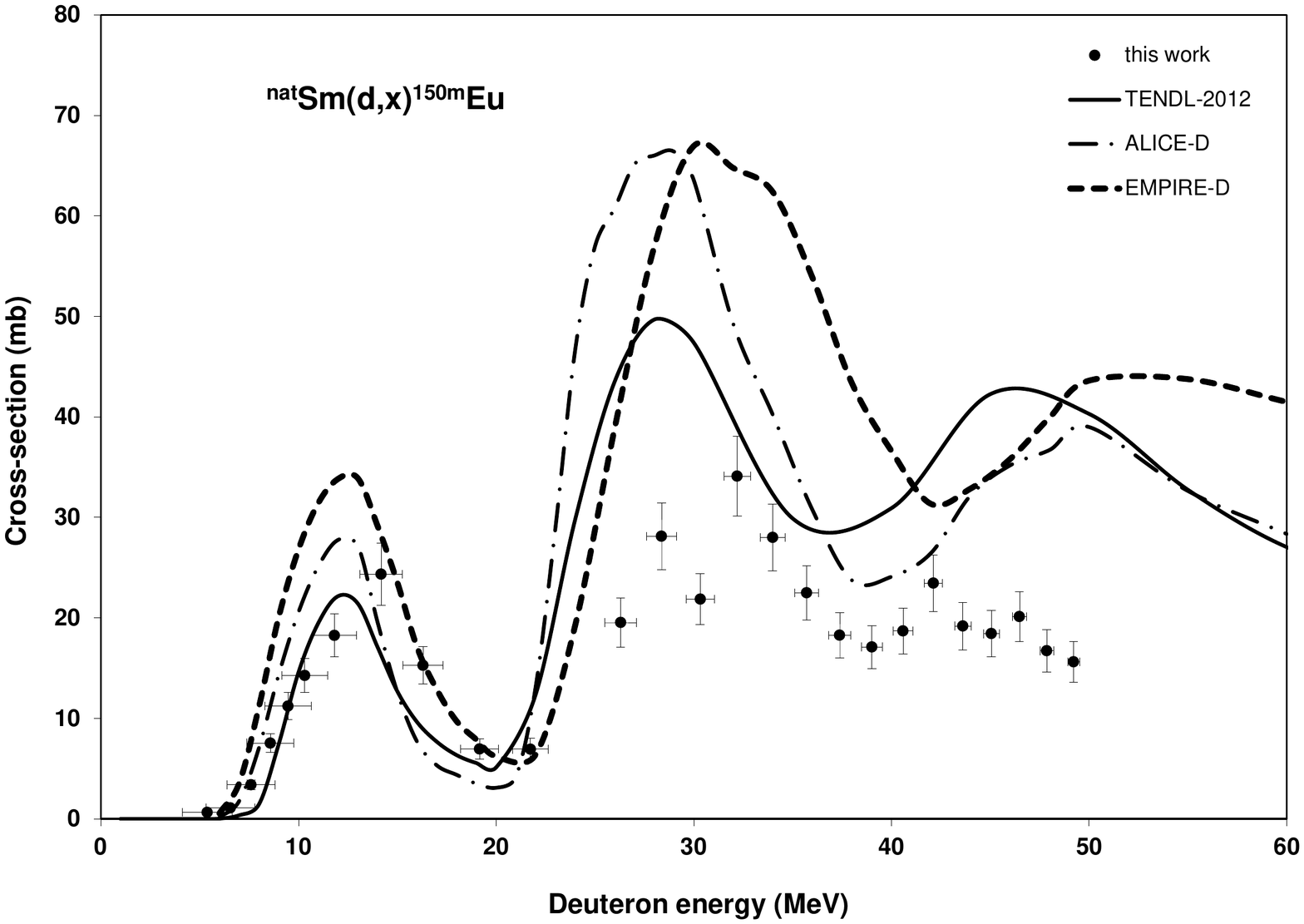}
\caption{Excitation functions of the $^{nat}$Sm(d,xn)$^{150m}$Eu reaction in comparison with results from model calculations}
\end{figure}

\vspace{0.4 cm}
\textbf{4.1.1.5	Production cross-sections of $^{150g}$Eu}\\

Only direct production has to be considered as practical, no contribution from the decay of $^{150m}$Eu exists (IT $<$ 10$^{-8}$, see previous paragraph). The independent cross-sections for direct production of  $^{154g}$Eu (GS, J$^\pi$ =  5$^-$, T$_{1/2}$ = 36.9 a,  $\varepsilon$: 100 \% ) are compared with the theory in Fig 7.  The three codes represent well the shape of the experimental data and the values of the maximum of the (d,n) reaction and TENDL-2012 also gives acceptable results at energies above 30 MeV.

\begin{figure}
\includegraphics[scale=0.3]{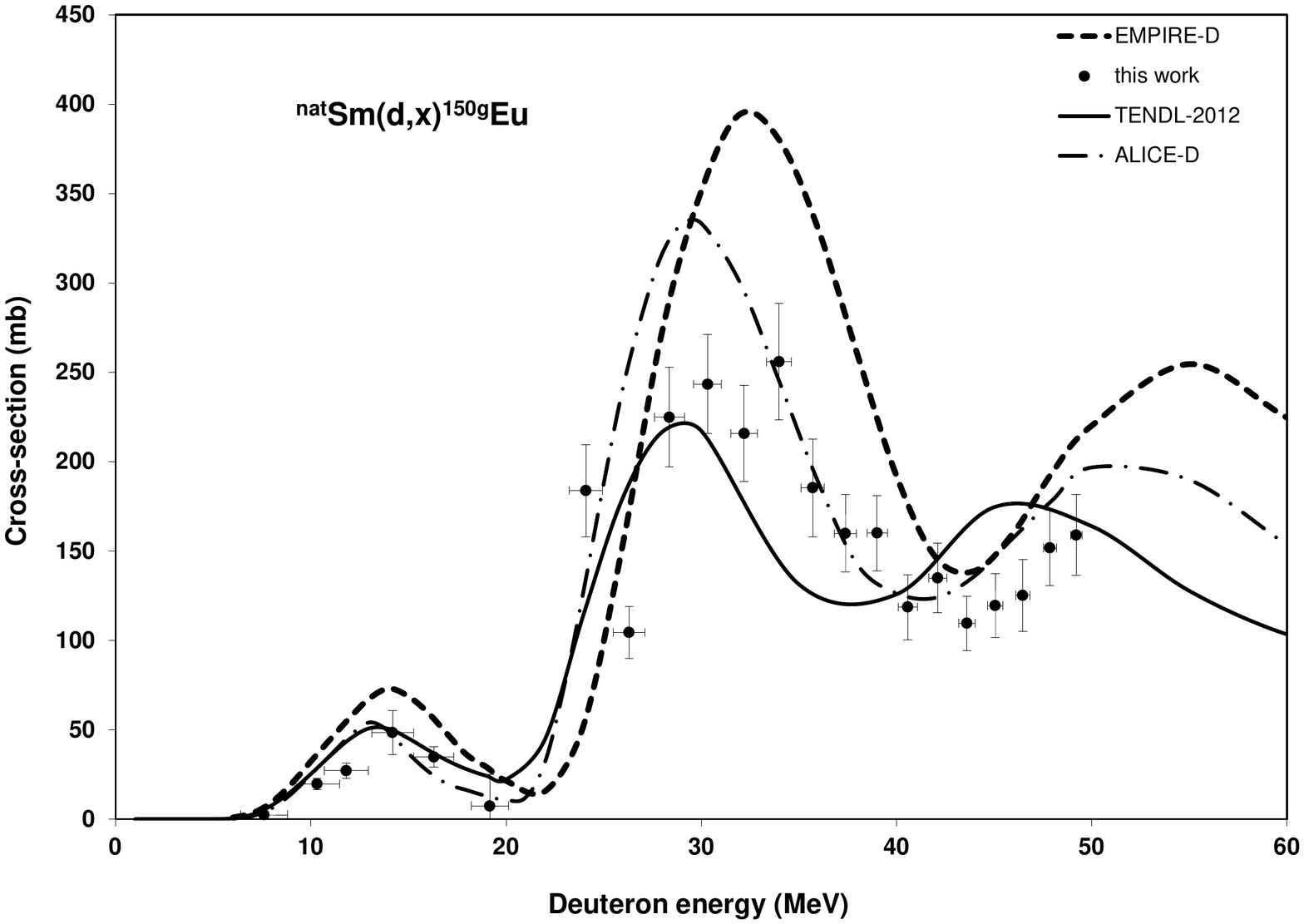}
\caption{Excitation functions of the $^{nat}$Sm(d,xn)$^{150g}$Eu reaction in comparison with results from model calculations}
\end{figure}

\vspace{0.4 cm}
\textbf{4.1.1.6	Production cross-sections of  $^{149}$Eu}\\

There are large disagreements between the results of the theoretic models and the experimental data for production of $^{149}$Eu (GS, J$^\pi$ =  5/2$^+$, T$_{1/2}$ = 93.1 d, $\varepsilon$: 100 \%) (Fig. 8). At low energy range the agreement with TENDL and ALICE-D is acceptable good, but the EMPIRE overestimation is significant. At the high energy range the shape of the ALICE-D and EMPIRE-D is similar to the experiment but the magnitude is much higher. The TENDL underestimates the peak for the (d,5n) reaction and shows a separate contribution for the (d,7n) reaction, lumped in the experiment and the two other codes.

\begin{figure}
\includegraphics[scale=0.3]{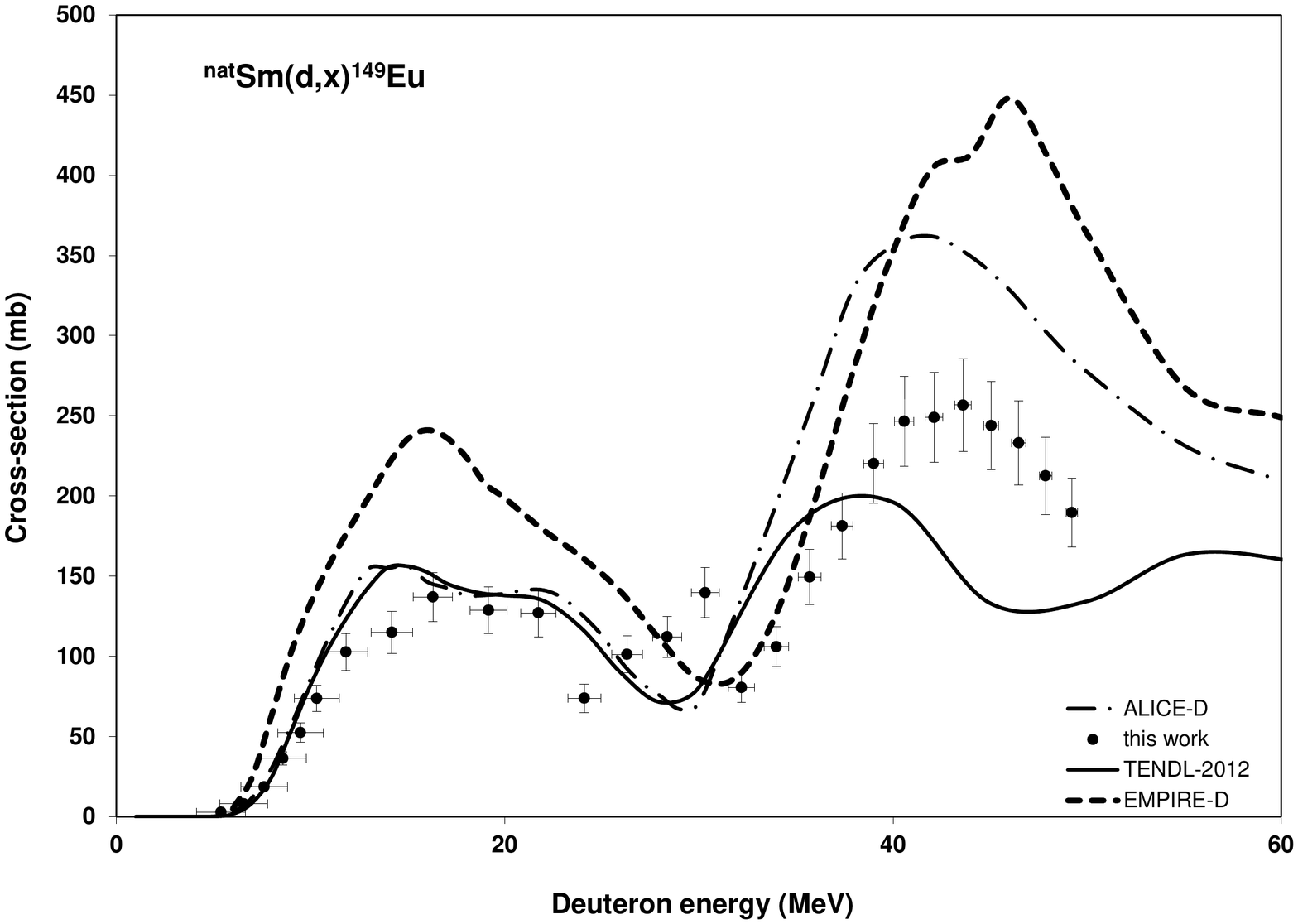}
\caption{Excitation functions of the $^{nat}$Sm(d,xn)$^{149}$Eu reaction in comparison with results from model calculations }
\end{figure}

\vspace{0.4 cm}
\textbf{4.1.1.7	Production cross-sections of  $^{148}$Eu}\\

The experimental and theoretical data for cross-sections of $^{148}$Eu (GS, J$^\pi$ = 5$^-$, T$_{1/2}$ = 54.5 d, $\varepsilon$: 100 \%, $\alpha$: 9.4*10$^{-7}$ \%) are compared in Fig. 9. In the energy domain considered reactions on 4 stable Sm isotopes can contribute below 35 MeV, which is reflected in the results of TENDL-2012 and ALICE-D. The shape for EMPIRE-D is much smoother especially at higher energies, where the reaction on $^{152}$Sm seems to be overestimated.  The agreement with our experimental results below the threshold of the (d,6n) reaction is  more or less acceptable, but large differences in energy dependency appear with the TENDL-2012 data above 35 MeV.

\begin{figure}
\includegraphics[scale=0.3]{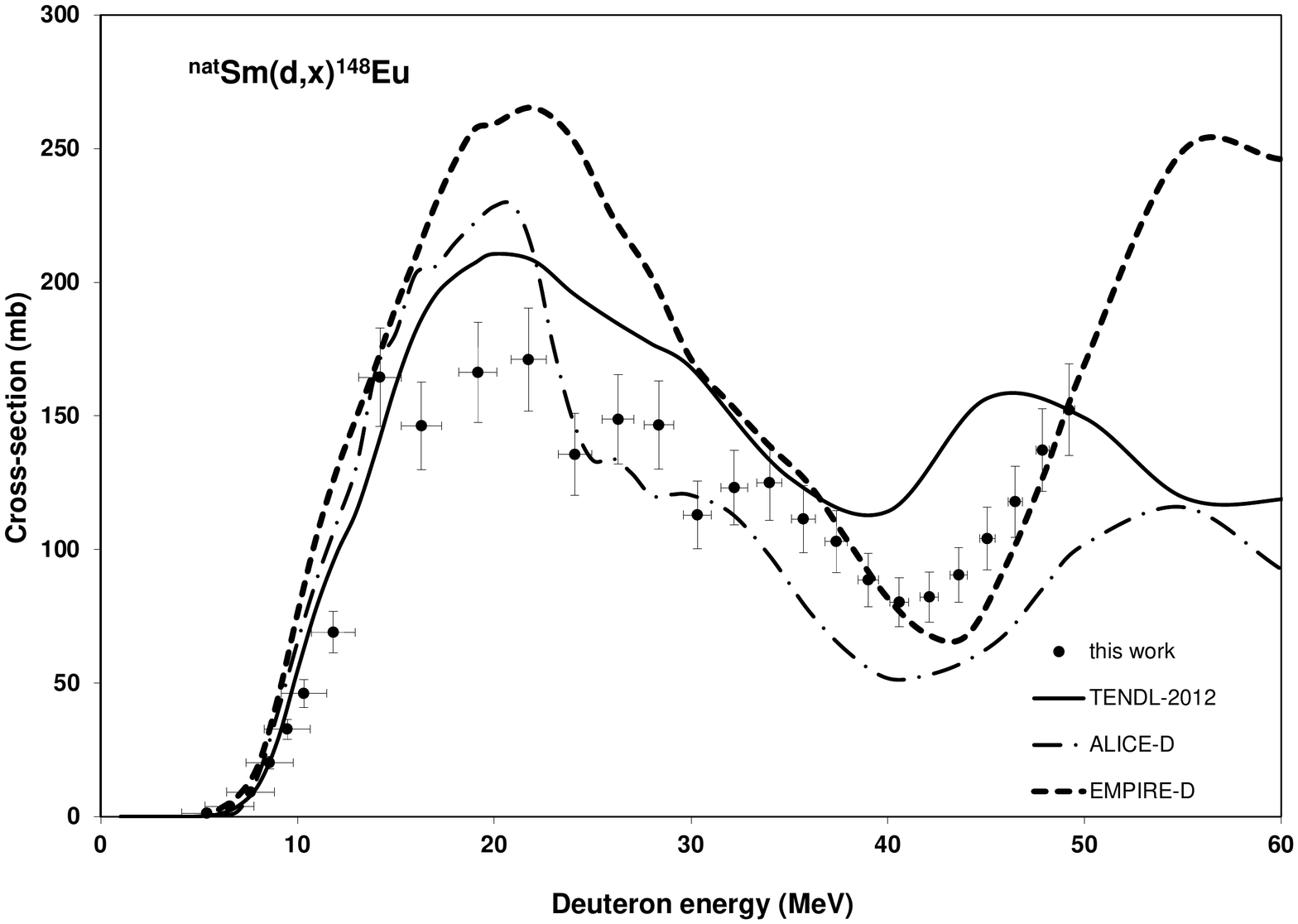}
\caption{Excitation functions of the $^{nat}$Sm(d,xn)$^{148}$Eu reaction in comparison with results from model calculations}
\end{figure}
\vspace{0.4 cm}
\textbf{4.1.1.8	Production cross-sections of  $^{147}$Eu}\\

There is acceptable agreement in the shape between the experiment and the theory for $^{147}$Eu (GS, J$^\pi$ = 5/2$^+$, T$_{1/2}$ = 24.1 d, $\varepsilon$: 99.9978 \%, $\alpha$: 0.0022 \%) cross-sections (Fig. 10). The experiment shows less detailed the contribution on the different stable Sm target isotopes.  The ALICE and EMPIRE codes overestimate the experimental data in the different peak regions and show energy shift with TENDL-2012 for the onset of the (d,7n) and (d,9n) reactions. 

\begin{figure}
\includegraphics[scale=0.3]{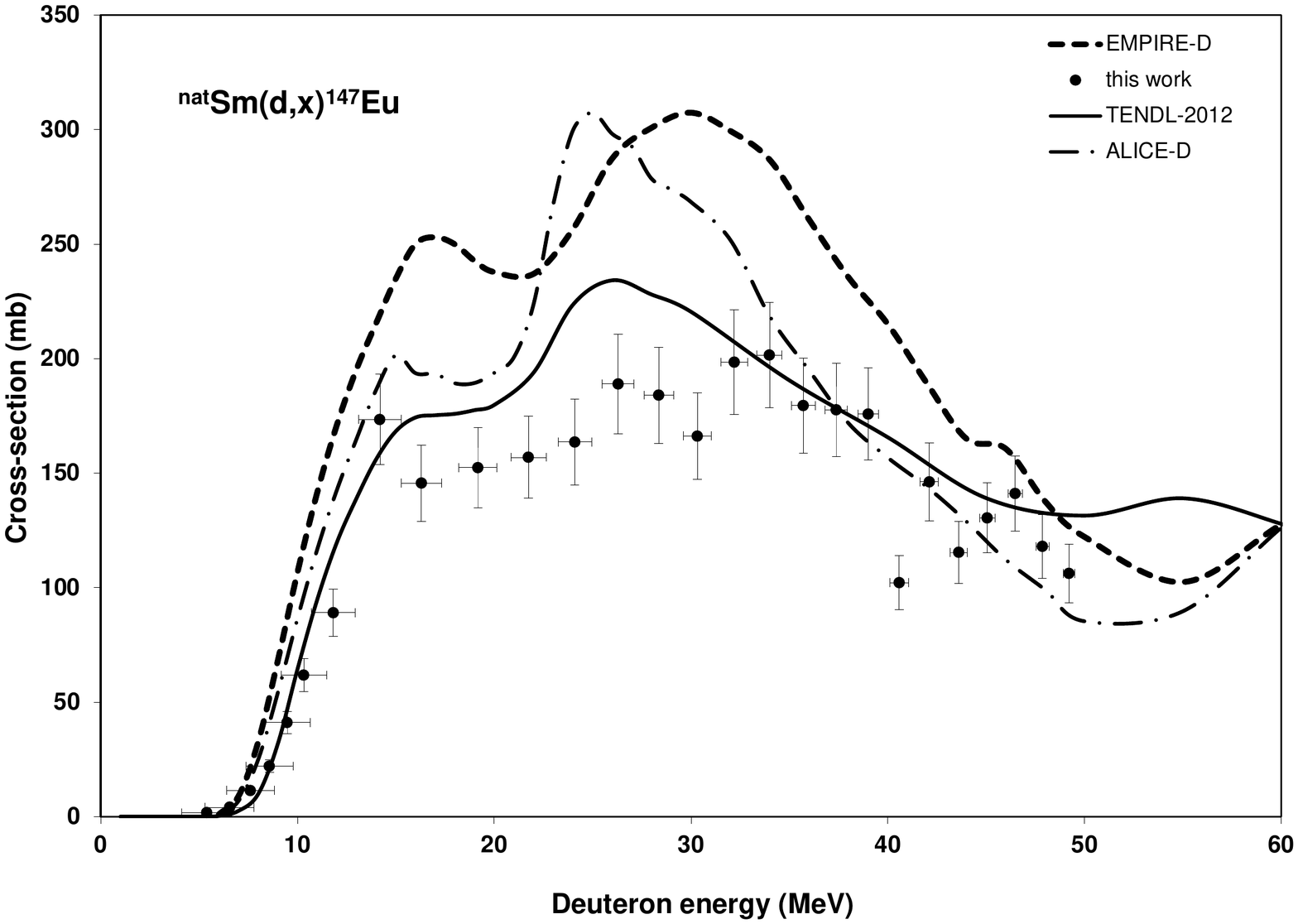}
\caption{Excitation functions of the $^{nat}$Sm(d,xn)$^{147}$Eu reaction in comparison with results from model calculations }
\end{figure}
\vspace{0.4 cm}
\textbf{4.1.1.9	Production cross-sections of  $^{146}$Eu}\\

The experimental results correspond well to the overall shape of the more detailed theoretical descriptions of the excitation function for $^{146}$Eu (GS, J$^\pi$ = 4$^-$, T$_{1/2}$ = 4.59 d,  $\varepsilon$:100 \%). For all models the overestimation in the whole energy range is significant (Fig. 11). 

\begin{figure}
\includegraphics[scale=0.3]{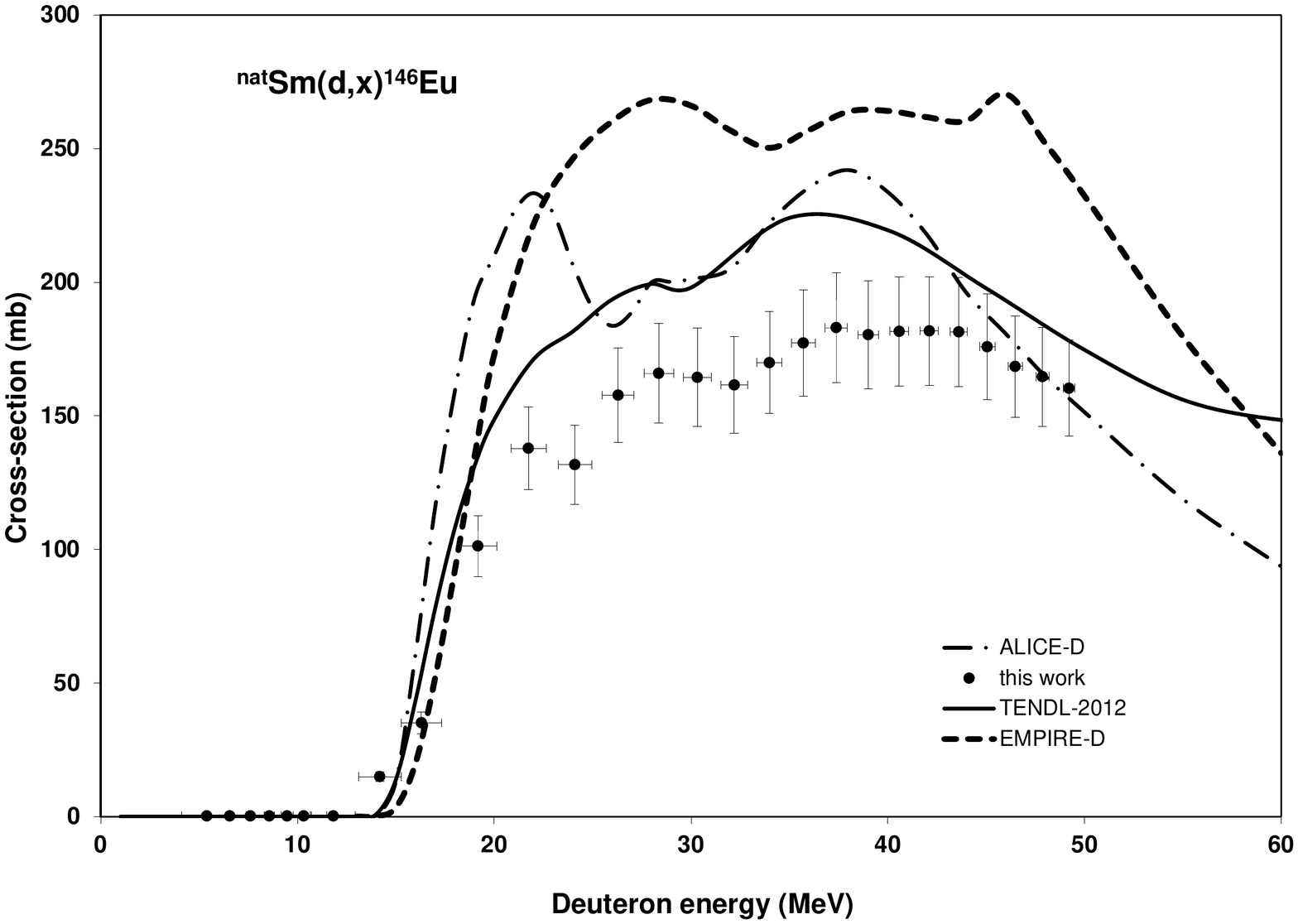}
\caption{Excitation functions of the $^{nat}$Sm(d,xn)$^{146}$Eu reaction in comparison with results from model calculations}
\end{figure}
\vspace{0.4 cm}
\textbf{4.1.1.10	Production cross-sections of  $^{145}$Eu}\\

The production of $^{145}$Eu (GS, J$^\pi$ =  5/2$^+$,  T$_{1/2}$ = 5.93 d,  $\varepsilon$: 100 \%) is in detail discussed in our previous work [4]. There are large overestimations in the magnitude in the theoretical results (Fig. 12) especially at energies above 30 MeV.

\begin{figure}
\includegraphics[scale=0.3]{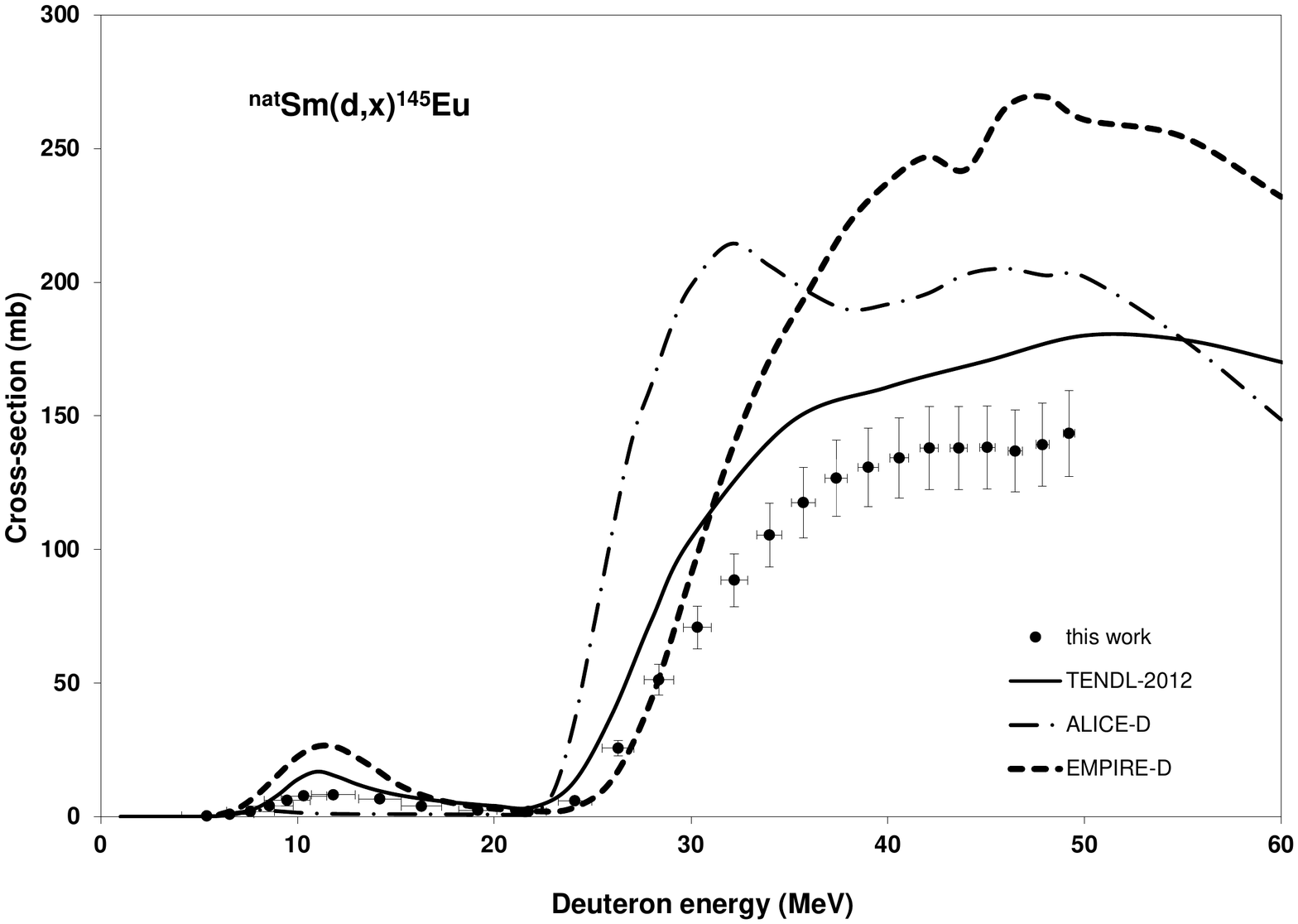}
\caption{Excitation functions of the $^{nat}$Sm(d,xn)$^{145}$Eu reaction in comparison with results from model calculations}
\end{figure}

\subsubsection{Production of radioisotopes of samarium}
\label{4.1.2}
 The $\gamma$-lines of the $^{155}$Sm due to the short half-life (22.3 min) comparing to the used cooling times were not detected.\\

\vspace{0.4 cm}
\textbf{4.1.2.1	Production cross-sections of  $^{153}$Sm}\\

The medically relevant radionuclide $^{153}$Sm (GS, J$^\pi$ = 3/2$^+$, T$_{1/2}$ = 46.50 h, $\beta^-$: 100 \%) is mainly produced directly through (d,p) and (d,p2n) reactions and through the decay of the short-lived $^{153}$Pm parent (GS, J$^\pi$ = 5/2$^-$, T$_{1/2}$ = 5.25 min,  $\beta^-$: 100 \%.) The presented cross-sections are cumulative, obtained from spectra measured after the decay of $^{153}$Pm (Fig. 13). More discussion on the production of $^{153}$Sm can be found in our previous work \cite{4}. There are large disagreements between the experiment and the theory both in shape and in magnitude confirming earlier findings on the different modeling of the breakup reaction in the different codes.

\begin{figure}
\includegraphics[scale=0.3]{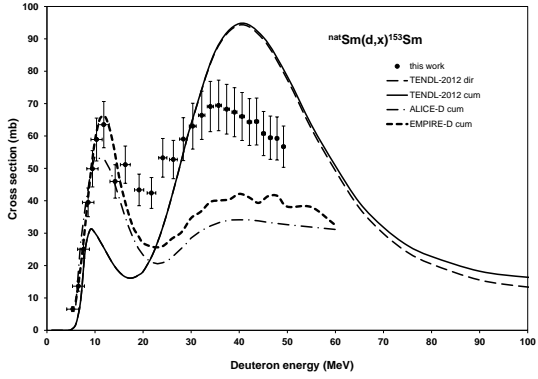}
\caption{Excitation functions of the $^{nat}$Sm(d,pxn)$^{153}$Sm reaction in comparison with results from model calculations}
\end{figure}
\vspace{0.4 cm}
\textbf{4.1.2.2	Production cross-sections of  $^{145}$Sm}\\

The cross-sections for cumulative production of $^{145}$Sm (GS,  J$^\pi$ = 7/2$^-$, T$_{1/2}$ = 340 d,  $\varepsilon$: 100 \%) are shown in Fig. 14. It includes the direct production through (d,pxn) reactions and the contribution from the decay of $^{145}$Eu (GS,  J$^\pi$ = 5/2$^+$,  T$_{1/2}$ = 5.93 d,  $\varepsilon$: 100 \%) The production is discussed in more detail in \cite{4}. The overestimation by the ALICE and EMPIRE is significant, cumulative TENDL calculation gives acceptable agreement. 

\begin{figure}
\includegraphics[scale=0.3]{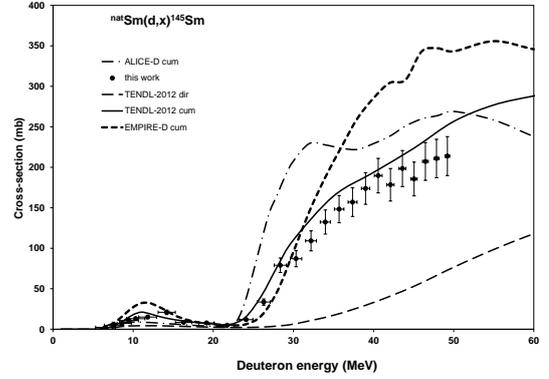}
\caption{Excitation functions of the $^{nat}$Sm(d,xn)$^{145}$Sm reaction in comparison with results from model calculations }
\end{figure}

\subsubsection{Production of radioisotopes of promethium}
\label{4.1.3}

\vspace{0.4 cm}
\textbf{4.1.3.1	Production cross-sections of  $^{151}$Pm}\\

The cross-section of the $^{151}$Pm (GS, J$^\pi$ =5/2$^+$, T$_{1/2}$ = 28.40 h, $\beta^-$: 100 \%) is practically independent, direct production cross-section through (d,2pn) reaction on $^{152}$Sm taking into account that the cross-section of the 151Nd (12.44 min) is negligible small (d,3pxn)  and the half-life of the $^{151}$Sm other parent isotope is very high (90 a). The agreement with the theoretical data is acceptable good (Fig. 15), especially with TENDL below 40 MeV. Over this energy the judgment is difficult because of the large scattering in the experimental points.

\begin{figure}
\includegraphics[scale=0.3]{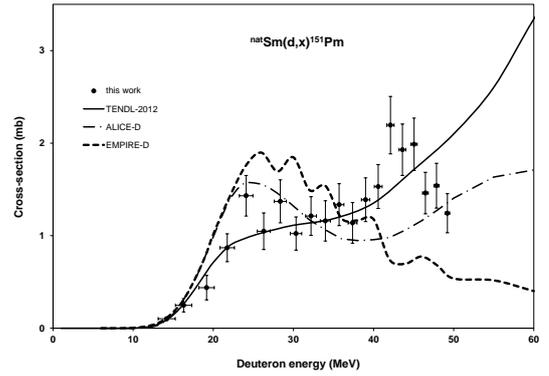}
\caption{Excitation functions of the $^{nat}$Sm(d,xn)$^{151}$Pm reaction in comparison with results from model calculations}
\end{figure}
\vspace{0.4 cm}
\textbf{4.1.3.2	Production cross-sections of $^{150}$Pm}\\

The radionuclide $^{150}$Pm (GS, J$^\pi$ = (1$^-$), T$_{1/2}$ = 2.68 h,  $\beta^-$) is a closed nuclei from the  point of view of decay of parents. It is hence produced only directly via (d,2pxn) reactions.  The measured cross-sections are higher than the theoretical predictions, especially in the lower energy region corresponding to a combination of $^{150}$Sm(d,2p) and $^{152}$Sm(d,$\alpha$) reactions  (Fig. 16).

\begin{figure}
\includegraphics[scale=0.3]{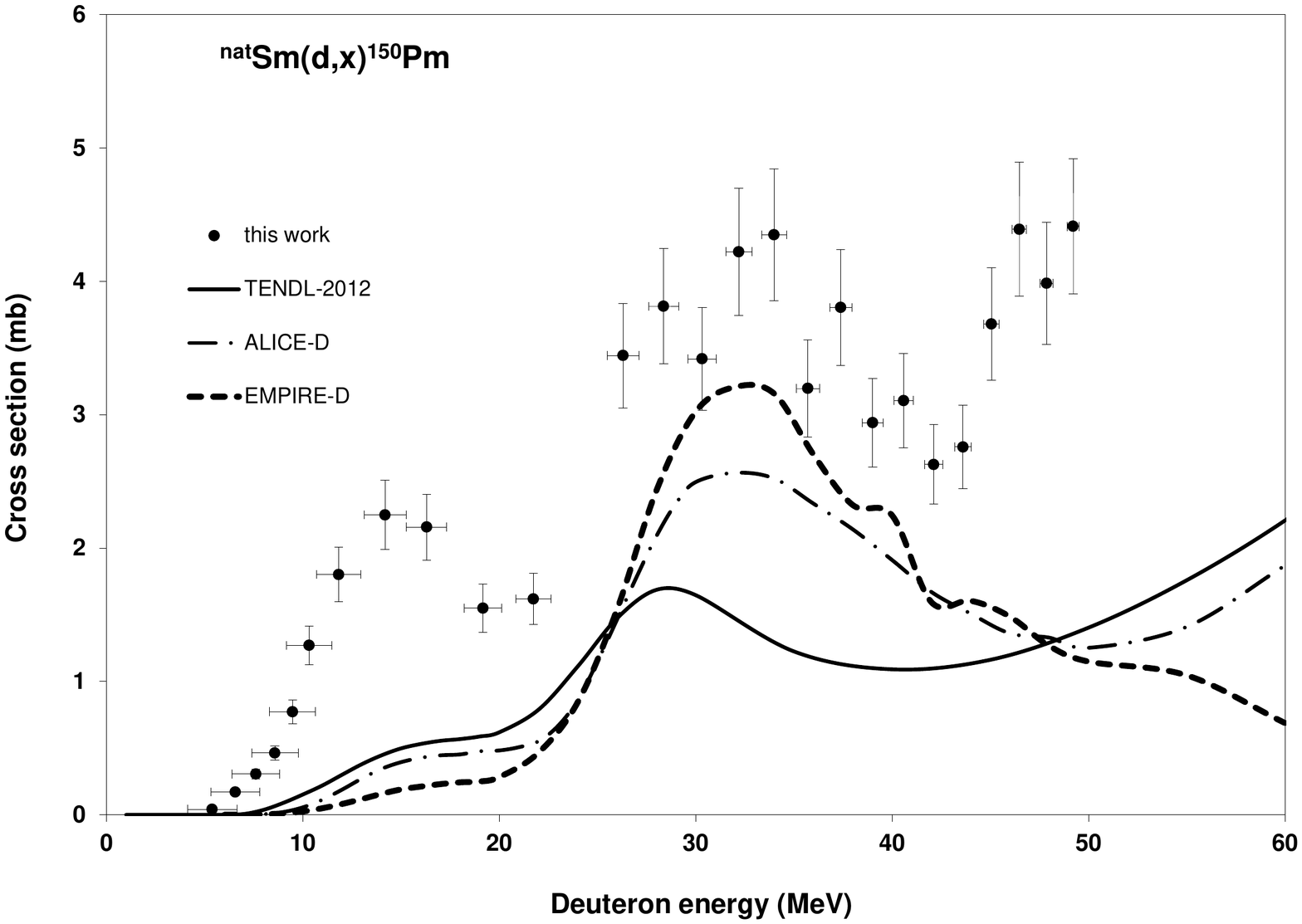}
\caption{Excitation functions of the $^{nat}$Sm(d,xn)$^{150}$Pm reaction in comparison with results from model calculations }
\end{figure}
\vspace{0.4 cm}
\textbf{4.1.3.3	Production cross-sections of  $^{149}$Pm}\\

Neglecting the small contribution from decay  149 Nd  (GS, J$^\pi$ =  5/2$^-$,T$_{1/2}$ = 1.728 h,   $\beta^-$: 100 \%) produced by (d,3pxn) reactions with low cross-sections, the $^{149}$Pm (GS,  J$^\pi$ = 7/2$^+$, 53.08 h, $\beta^-$: 100 \%) is produced only directly. We have only few experimental points due to the low cross-sections. Only in the measurements of a few target foils at higher energies (long measurements) was the statistics sufficient to calculate reliable cross-sections (Fig 17). According the few measured point the TENDL-2012 estimation seems to be acceptable.

\begin{figure}
\includegraphics[scale=0.3]{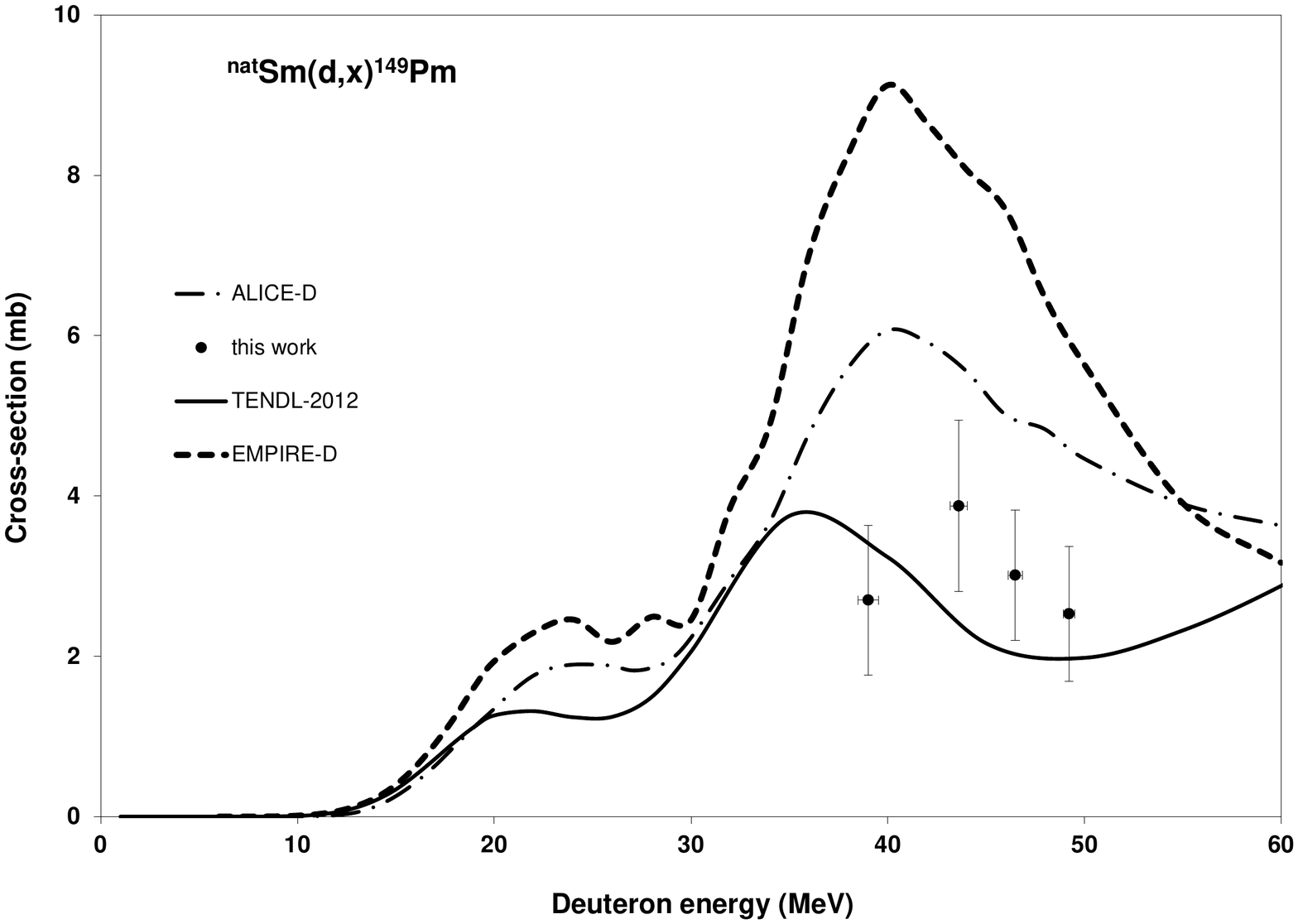}
\caption{Excitation functions of the $^{nat}$Sm(d,xn)$^{149}$Pm reaction in comparison with results from model calculations}
\end{figure}
\vspace{0.4 cm}
\textbf{4.1.3.4	Production cross-sections of  $^{148}$Pm}\\

The theoretical cross-sections of the directly produced $^{148}$Pm (MS, E(level)= 137.93 keV,  J$^\pi$ = 5$^-$,6$^-$, T$_{1/2}$ =  41.29 d, IT: 4.2 6 \%, $\beta^-$: 95.8 \%) are shown in Fig 18. The $^{148}$Pm (T$_{1/2}$  = 41.29 d) and the simultaneously produced $^{148}$Eu (T$_{1/2}$ = 54.5 d)  have similar half-lives and practically all $\gamma$-lines of $^{148}$Pm are common with $\gamma$-lines of $^{148}$Eu. The $^{148}$Eu has significantly higher cross-sections therefore the separation of the gamma intensities can be done with very large uncertainties. The measurements show much larger values than any of the theoretical calculations up to 30 MeV, and only the EMPIRE estimation is acceptable between 30 and 40 MeV and that of ALICE in the 45-50 MeV region.

\begin{figure}
\includegraphics[scale=0.3]{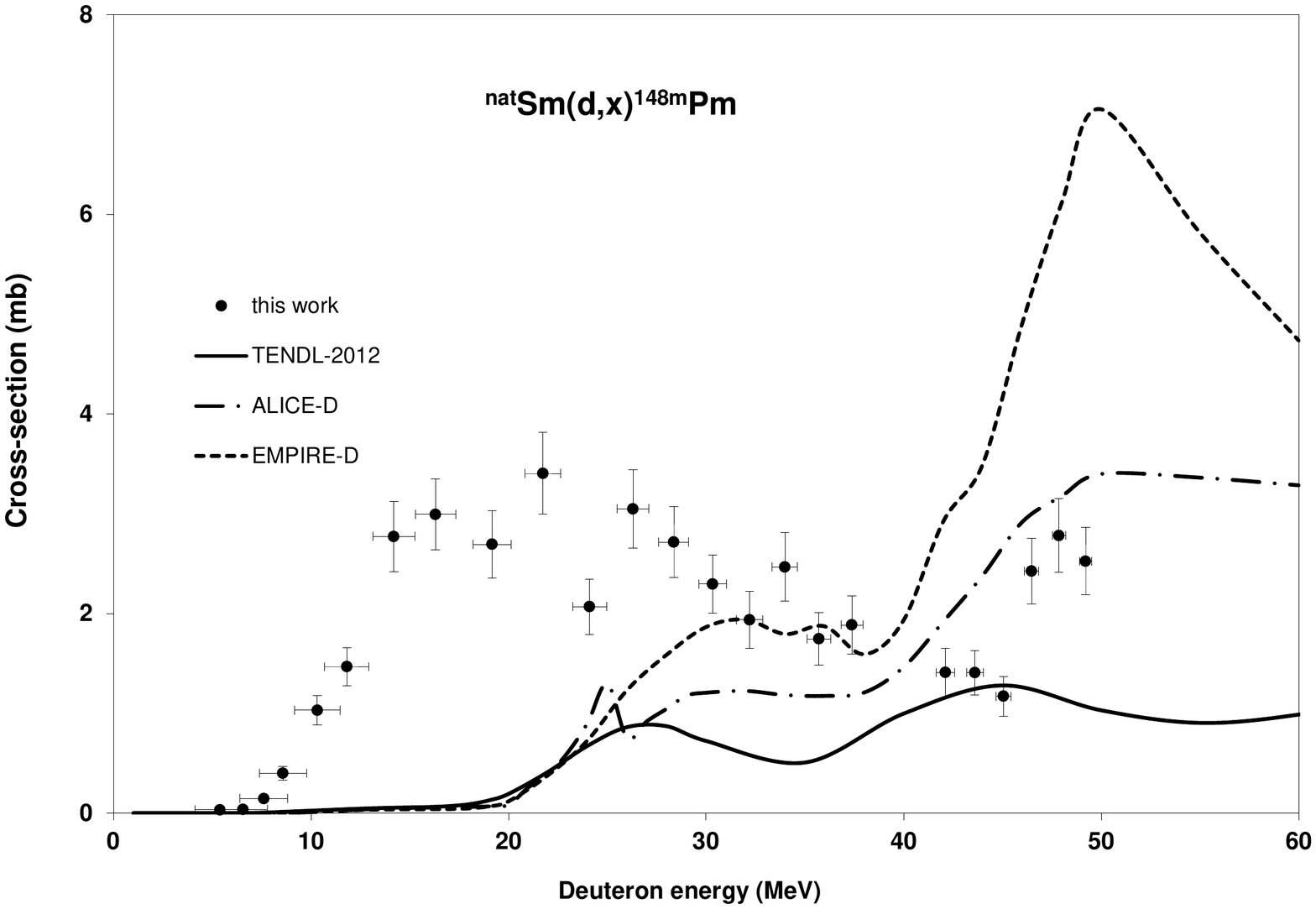}
\caption{Excitation functions of the $^{nat}$Sm(d,xn)$^{148}$Pm reaction in comparison with results from model calculations}
\end{figure}
\vspace{0.4 cm}
\textbf{4.1.3.5	Production cross-sections of $^{148g}$Pm}\\

The ground state of $^{148g}$Pm (GS,  J$^\pi$= 1$^-$,  T$_{1/2}$ = 5.368 d,  $\beta^-$: 100 \%) is produced directly and from the weak isomeric decay  of the longer-lived metastable state (41.29 d, IT: 4.2 \%,). We were not able to identify in any of our spectra the only strong independent $\gamma$-line (1465.12 keV, 22.2 \%) of $^{148g}$Pm. In principle it is possible to obtain data for direct production of $^{148g}$Pm from an analysis of the time dependence of the count rate for 550 keV and 915 keV $\gamma$-lines common between the ground state, the isomeric state and the $^{148}$Eu, but due to the low cross-section we came to the conclusion not to do an unreliable separation. The theoretical results are shown in Fig. 19. 

\begin{figure}
\includegraphics[scale=0.3]{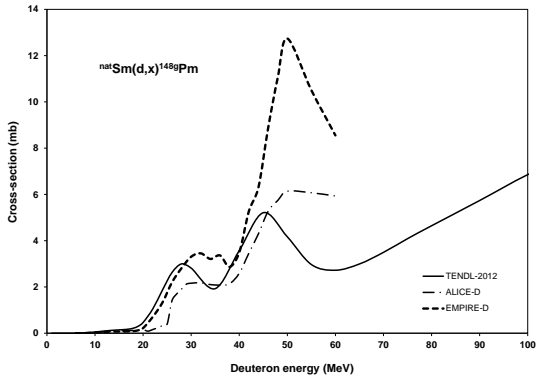}
\caption{Excitation functions of the $^{nat}$Sm(d,xn)$^{148g}$Pm  reaction in comparison with results from model calculations}
\end{figure}
\vspace{0.4 cm}
\textbf{4.1.3.6	Production cross-sections of $^{147}$Pm}\\

The $^{147}$Pm was not quantified due to its long half-life (2.6234 a), the low abundance of 121 keV $\gamma$-line (0.00285 \%) and the overlapping $\gamma$-lines from long-lived Eu radio-products.

\vspace{0.4 cm}
\textbf{4.1.3.7	Production cross-sections of $^{146}$Pm}

The $^{146}$Pm  (GS, J$^\pi$ = 3$^-$, T$_{1/2}$ = 5.53 a, $\varepsilon$: 66.0 \%) can only be produced by (d,2pxn) reactions directly. The excitation function is shown in Fig. 20. The experimental data are higher (but they are scattered and have large uncertainties) compared to the results of the model calculations.

\begin{figure}
\includegraphics[scale=0.3]{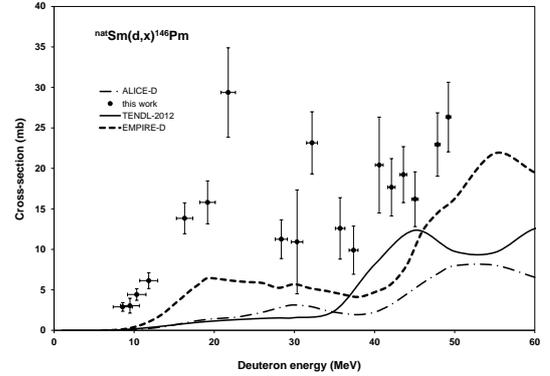}
\caption{Excitation functions of the $^{nat}$Sm(d,xn)$^{146}$Pm reaction in comparison with results from model calculations}
\end{figure}
\vspace{0.4 cm}
\textbf{4.1.3.8	Production cross-sections of  $^{145}$Pm}\\

The $\gamma$-lines of the $^{145}$Pm were not detected due to the long half-life (17.7 a) and it has only low energy low intensity $\gamma$-lines (and complex X-ray spectra).

\vspace{0.4 cm}
\textbf{4.1.3.9	Production cross-sections of  $^{144}$Pm}\\

The experimental and theoretical data for production of the radioisotope $^{144}$Pm (GS, J$^\pi$ = 5$^-$, T$_{1/2}$ = 363 d, $\varepsilon$: 100 \%) are shown in Fig. 21.  As $^{144}$Pm is a closed nuclei there is no parent decay possible. The experimental data are systematically lower compared to ALICE-D and EMPIRE-D and a little higher than the TALYS(TENDL) results.

\begin{figure}
\includegraphics[scale=0.3]{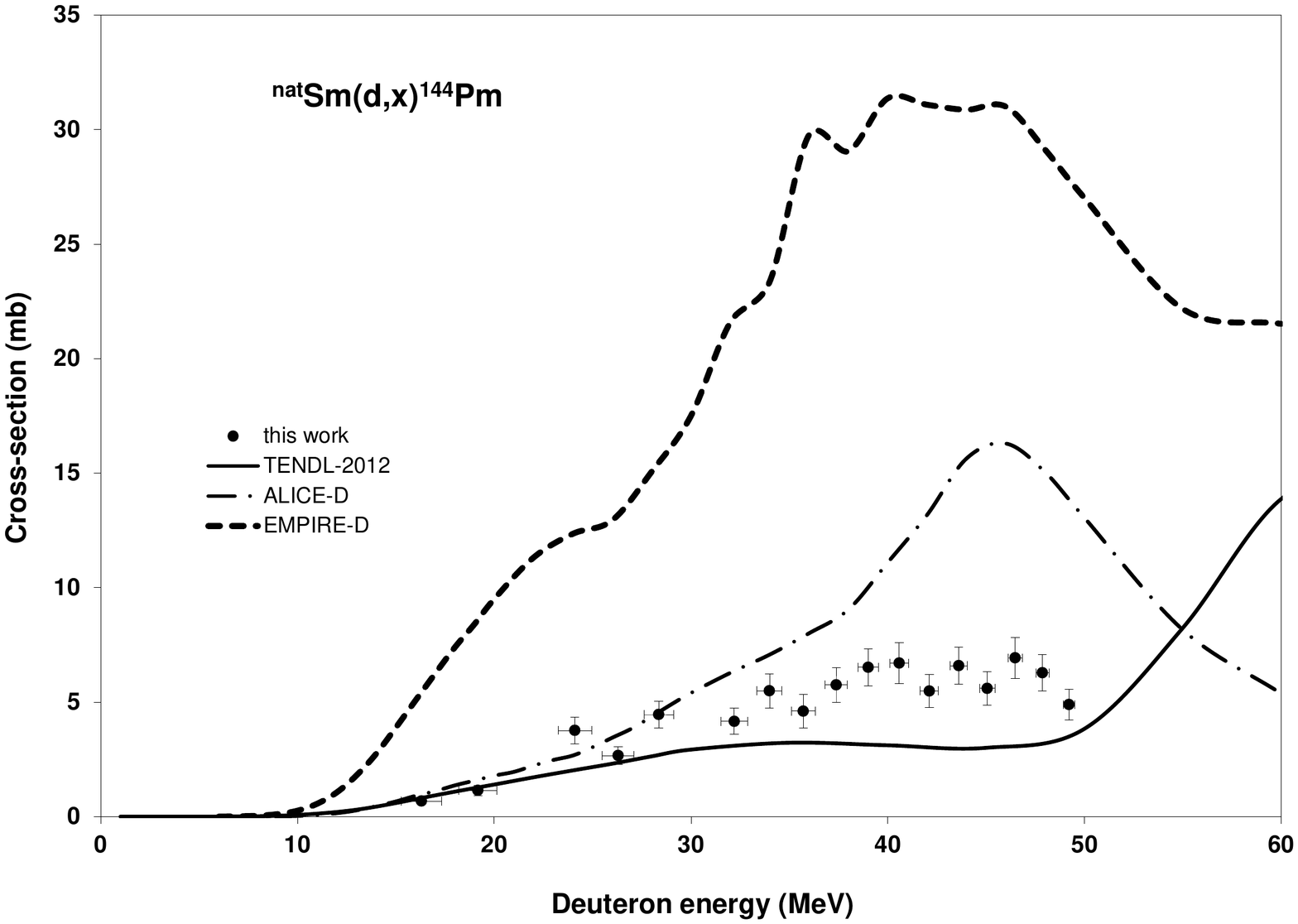}
\caption{Excitation functions of the $^{nat}$Sm(d,xn)$^{144}$Pm reaction in comparison with results from model calculations}
\end{figure}
\vspace{0.4 cm}
\textbf{4.1.3.10	Production cross-sections of  $^{143}$Pm}\\

Our experimental cross-sections of the $^{143}$Pm (GS, J$^\pi$ = 5/2$^+$,  265 d, $\varepsilon$: 100 \%)  contain the direct production and the total contribution from the short lived parent $^{143}$Sm  states that we could not asses in this study:  from $^{143m}$Sm (MS, E(level) = 753.9916 keV, J$^\pi$ = 11/2$^-$, T$_{1/2}$ = 66 s,  $\varepsilon$: 0.24 \%, IT: 99.76 \%) and $^{143g}$Sm (GS, J$^\pi$ = 3/2$^+$,  T$_{1/2}$ = 8.75 min,  $\varepsilon$: 100 \%). As it is shown in Fig. 22 the ALICE-D prediction is very close to the experimental data above 30 MeV, but under this energy EMPIRE-D gives much better results. The underestimation of TENDL-2012 is significant. It has to be mentioned that the degree of detailed representation of the different contributing reactions is quite different for the three codes.

\begin{figure}
\includegraphics[scale=0.3]{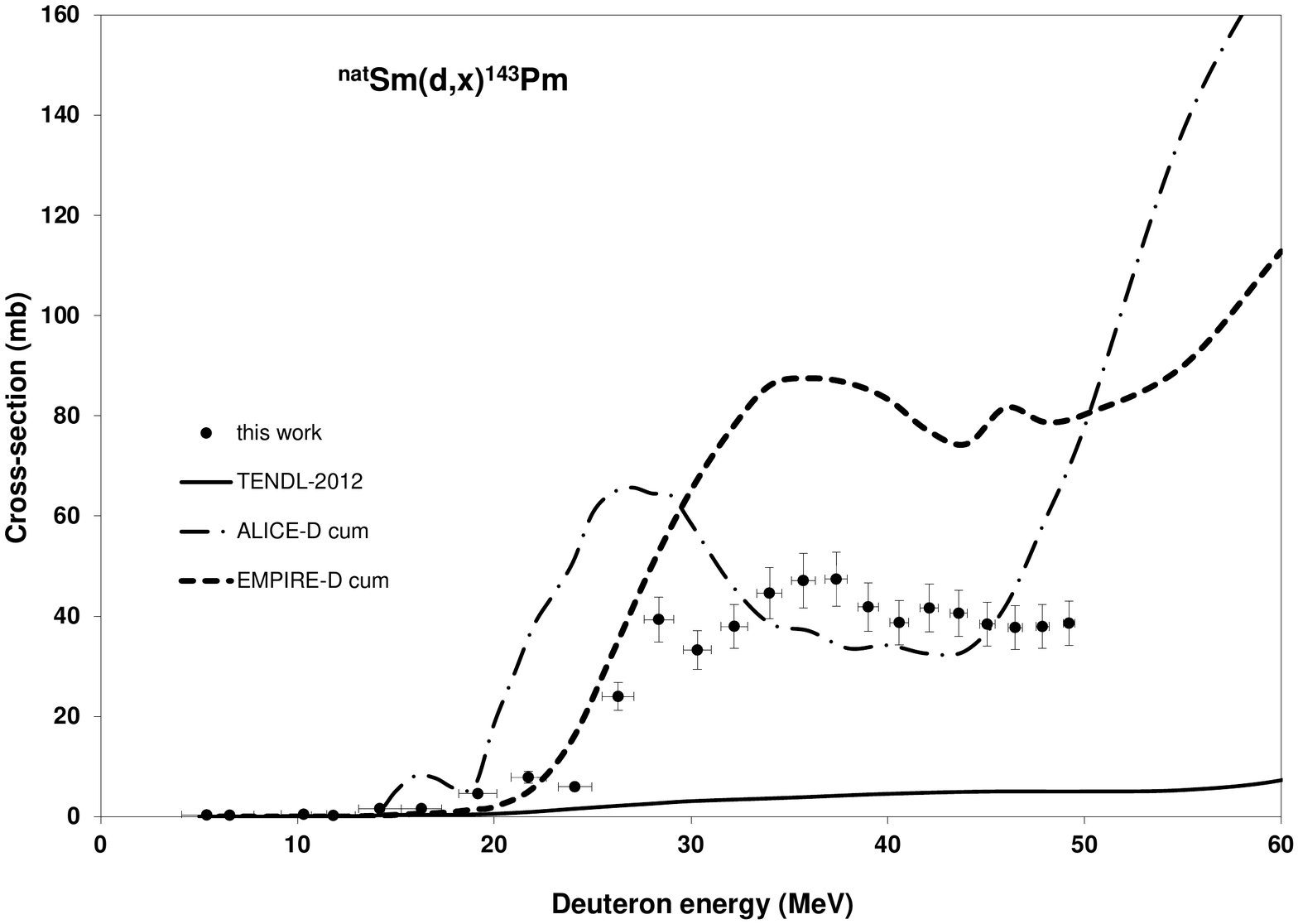}
\caption{Excitation functions of the $^{nat}$Sm(d,xn)$^{143}$Pm reaction in comparison with results from model calculations}
\end{figure}

\begin{table*}[t]
\tiny
\caption{Measured cross-sections of the ${}^{155,154,152m2,152m1,152g,150m,150g,149}$Eu radioisotopes}
\centering
\begin{center}
\begin{tabular}{|c|c|c|c|c|c|c|c|c|c|c|c|c|c|c|c|c|c|} \hline 
\multicolumn{2}{|c|}{\textbf{Energy\newline }(MeV) } & \multicolumn{2}{|c|}{\textbf{${}^{155}$Eu \newline }$\sigma\pm\Delta\sigma$(mb)} & \multicolumn{2}{|c|}{\textbf{${}^{154}$Eu \newline }$\sigma\pm\Delta\sigma$(mb)} & \multicolumn{2}{|c|}{\textbf{${}^{152m2}$Eu \newline }$\sigma\pm\Delta\sigma$(mb)} & \multicolumn{2}{|c|}{\textbf{${}^{152m1}$Eu \newline }$\sigma\pm\Delta\sigma$(mb)} & \multicolumn{2}{|p{0.7in}|}{\textbf{${}^{152g}$Eu \newline }$\sigma\pm\Delta\sigma$(mb)} & \multicolumn{2}{|p{0.7in}|}{\textbf{${}^{150m}$Eu \newline }$\sigma\pm\Delta\sigma$(mb)} & \multicolumn{2}{|c|}{\textbf{${}^{150g}$Eu \newline }$\sigma\pm\Delta\sigma$mb)} & \multicolumn{2}{|c|}{\textbf{${}^{149}$Eu \newline }$\sigma\pm\Delta\sigma$(mb)} \\ \hline 
49.2 & 0.3 & 6.4 & 1.1 & ~ &  & 5.0 & 0.6 & 6.4 & 0.8 & ~ &  & 15.6 & 2.0 & 158.9 & 22.6 & 189.6 & 21.5 \\ \hline 
47.9 & 0.3 & ~ &  & ~ &  & 4.7 & 0.5 & 7.7 & 0.9 & 41.9 & 15.1 & 16.7 & 2.1 & 151.8 & 21.0 & 212.6 & 24.1 \\ \hline 
46.5 & 0.4 & ~ &  & ~ &  & 5.1 & 0.6 & 7.9 & 0.9 & 76.1 & 20.2 & 20.1 & 2.5 & 125.2 & 20.1 & 233.1 & 26.4 \\ \hline 
45.1 & 0.4 & ~ &  & ~ &  & 5.5 & 0.6 & 6.9 & 0.8 & 59.4 & 15.4 & 18.4 & 2.3 & 119.5 & 17.9 & 243.9 & 27.5 \\ \hline 
43.6 & 0.4 & ~ &  & ~ &  & 5.7 & 0.6 & 8.3 & 1.0 & 82.5 & 15.1 & 19.2 & 2.4 & 109.6 & 15.2 & 256.6 & 28.9 \\ \hline 
42.1 & 0.5 & 10.7 & 2.8 & ~ &  & 6.3 & 0.7 & 8.5 & 1.0 & 132.7 & 21.4 & 23.4 & 2.8 & 134.9 & 19.4 & 249.0 & 28.1 \\ \hline 
40.6 & 0.5 & 10.3 & 3.3 & ~ &  & 7.2 & 0.8 & 10.2 & 1.2 & 159.8 & 25.9 & 18.7 & 2.3 & 118.7 & 18.2 & 246.5 & 27.9 \\ \hline 
39.0 & 0.5 & 10.8 & 3.0 & ~ &  & 8.2 & 0.9 & 10.7 & 1.2 & 148.3 & 22.2 & 17.1 & 2.1 & 160.1 & 21.0 & 220.2 & 24.9 \\ \hline 
37.4 & 0.6 & ~ &  & ~ &  & 10.5 & 1.2 & 12.0 & 1.4 & 144.6 & 23.5 & 18.3 & 2.2 & 159.8 & 21.7 & 181.4 & 20.6 \\ \hline 
35.7 & 0.6 & ~ &  & ~ &  & 13.4 & 1.5 & 13.0 & 1.5 & 116.0 & 24.8 & 22.5 & 2.7 & 185.4 & 27.3 & 149.4 & 17.2 \\ \hline 
34.0 & 0.6 & 10.0 & 3.7 & ~ &  & 15.1 & 1.7 & 14.1 & 1.6 & 186.4 & 27.9 & 28.0 & 3.3 & 255.9 & 32.5 & 106.0 & 12.3 \\ \hline 
32.2 & 0.7 & 17.5 & 3.5 & ~ &  & 16.1 & 1.8 & 17.3 & 2.0 & 211.5 & 27.9 & 34.1 & 4.0 & 215.8 & 26.9 & 80.5 & 9.4 \\ \hline 
30.3 & 0.7 & 11.5 & 1.8 & ~ &  & 13.0 & 1.5 & 19.2 & 2.2 & 174.5 & 20.6 & 21.9 & 2.5 & 243.3 & 27.7 & 139.7 & 15.7 \\ \hline 
28.4 & 0.8 & 9.6 & 2.8 & ~ &  & 13.6 & 1.5 & 20.0 & 2.3 & 104.4 & 54.3 & 28.1 & 3.3 & 224.9 & 27.9 & 112.1 & 12.8 \\ \hline 
26.3 & 0.8 & 24.7 & 3.8 & 58.2 & 8.5 & 10.2 & 1.1 & 18.1 & 2.1 & 170.3 & 22.7 & 19.5 & 2.4 & 104.4 & 14.4 & 101.2 & 11.5 \\ \hline 
24.1 & 0.9 & 12.4 & 8.0 & ~ &  & ~ &  & ~ &  & 151.2 & 28.1 & ~ &  & 183.8 & 25.7 & 73.8 & 8.9 \\ \hline 
21.7 & 0.9 & 29.1 & 5.7 & ~ &  & 3.0 & 0.4 & 10.8 & 1.2 & ~ &  & 6.9 & 1.1 & ~ &  & 127.0 & 14.7 \\ \hline 
19.2 & 1.0 & 39.3 & 4.9 & 94.6 & 11.9 & 3.2 & 0.4 & 14.0 & 1.6 & 71.9 & 14.5 & 6.9 & 1.0 & 7.4 & 17.8 & 128.7 & 14.6 \\ \hline 
16.3 & 1.0 & 48.0 & 5.5 & 115.1 & 13.3 & 5.1 & 0.6 & 24.5 & 2.8 & 116.1 & 14.5 & 15.3 & 1.8 & 34.8 & 5.7 & 137.0 & 15.4 \\ \hline 
14.2 & 1.1 & 36.7 & 5.1 & 171.9 & 22.2 & 9.7 & 1.1 & 25.1 & 2.9 & 133.7 & 20.5 & 24.3 & 3.1 & 48.4 & 12.2 & 114.9 & 13.2 \\ \hline 
11.8 & 1.1 & 59.2 & 6.7 & 156.0 & 17.7 & 2.9 & 0.3 & 42.1 & 4.8 & 142.7 & 16.7 & 18.3 & 2.1 & 27.1 & 4.2 & 102.7 & 11.6 \\ \hline 
10.3 & 1.2 & 58.1 & 6.6 & 118.3 & 13.4 & 1.3 & 0.2 & 34.3 & 3.9 & 89.2 & 10.5 & 14.3 & 1.7 & 19.7 & 3.0 & 73.8 & 8.3 \\ \hline 
9.5 & 1.2 & 48.8 & 5.8 & 95.3 & 11.9 & 0.7 & 0.1 & 25.3 & 2.9 & 55.8 & 9.6 & 11.2 & 1.3 & ~ &  & 52.4 & 6.0 \\ \hline 
8.6 & 1.2 & 39.4 & 4.5 & 51.2 & 6.0 & 0.4 & 0.0 & 18.1 & 2.0 & 29.8 & 4.3 & 7.5 & 0.9 & ~ &  & 36.3 & 4.1 \\ \hline 
7.6 & 1.2 & 24.8 & 2.8 & 27.9 & 3.4 & 0.2 & 0.0 & 8.4 & 1.0 & 13.0 & 2.2 & 3.4 & 0.5 & 2.4 & 0.9 & 18.6 & 2.1 \\ \hline 
6.6 & 1.2 & 12.7 & 1.5 & 10.2 & 1.5 & 0.1 & 0.0 & 3.4 & 0.4 & 9.8 & 1.9 & 1.1 & 0.2 & ~ &  & 8.0 & 0.9 \\ \hline 
5.4 & 1.3 & 5.3 & 0.6 & ~ &  & ~ &  & 0.9 & 0.1 & ~ &  & 0.6 & 0.1 & ~ &  & 2.8 & 0.3 \\ \hline 
\end{tabular}
\end{center}
\end{table*}

\begin{table*}[t]
\tiny
\caption{Measured cross-sections of the ${}^{148,147,146}$Eu and ${}^{151,150,149,148m,146}$Pm radioisotopes}
\centering
\begin{center}
\begin{tabular}{|c|c|c|c|c|c|c|c|c|c|c|c|c|c|c|c|c|c|} \hline 
\multicolumn{2}{|c|}{\textbf{Energy\newline }(MeV) } & \multicolumn{2}{|c|}{\textbf{${}^{148}$Eu \newline }$\sigma\pm\Delta\sigma$(mb)} & \multicolumn{2}{|c|}{\textbf{${}^{147}$Eu \newline }$\sigma\pm\Delta\sigma$(mb)} & \multicolumn{2}{|c|}{\textbf{${}^{146}$Eu \newline }$\sigma\pm\Delta\sigma$(mb)} & \multicolumn{2}{|c|}{\textbf{${}^{151}$Pm \newline }$\sigma\pm\Delta\sigma$(mb)} & \multicolumn{2}{|c|}{\textbf{${}^{150}$Pm \newline }$\sigma\pm\Delta\sigma$(mb)} & \multicolumn{2}{|c|}{\textbf{${}^{149}$Pm \newline }$\sigma\pm\Delta\sigma$(mb)} & \multicolumn{2}{|c|}{\textbf{${}^{148m}$Pm \newline }$\sigma\pm\Delta\sigma$(mb)} & \multicolumn{2}{|c|}{\textbf{${}^{146}$Pm \newline }$\sigma\pm\Delta\sigma$(mb)} \\ \hline 
49.2 & 0.3 & 152.2 & 17.1 & 106.2 & 12.7 & 160.3 & 18.0 & 1.2 & 0.2 & 4.4 & 0.5 & 2.5 & 0.8 & 2.5 & 0.3 & 26.3 & 4.3 \\ \hline 
47.9 & 0.3 & 137.2 & 15.4 & 118.0 & 13.9 & 164.6 & 18.5 & 1.5 & 0.2 & 4.0 & 0.5 & ~ &  & 2.8 & 0.4 & 22.9 & 3.9 \\ \hline 
46.5 & 0.4 & 117.8 & 13.3 & 141.1 & 16.3 & 168.4 & 18.9 & 1.5 & 0.2 & 4.4 & 0.5 & 3.0 & 0.8 & 2.4 & 0.3 & ~ &  \\ \hline 
45.1 & 0.4 & 104.1 & 11.7 & 130.4 & 15.3 & 175.8 & 19.7 & 2.0 & 0.3 & 3.7 & 0.4 & ~ &  & 1.2 & 0.2 & 16.2 & 3.4 \\ \hline 
43.6 & 0.4 & 90.4 & 10.2 & 115.4 & 13.6 & 181.4 & 20.4 & 1.9 & 0.3 & 2.8 & 0.3 & 3.9 & 1.1 & 1.4 & 0.2 & 19.2 & 3.5 \\ \hline 
42.1 & 0.5 & 82.2 & 9.3 & 146.1 & 17.0 & 181.7 & 20.4 & 2.2 & 0.3 & 2.6 & 0.3 & ~ &  & 1.4 & 0.2 & 17.7 & 3.5 \\ \hline 
40.6 & 0.5 & 80.2 & 9.1 & 102.1 & 11.9 & 181.6 & 20.4 & 1.5 & 0.2 & 3.1 & 0.4 & ~ &  & ~ &  & 20.4 & 5.9 \\ \hline 
39.0 & 0.5 & 88.6 & 10.0 & 175.7 & 20.2 & 180.3 & 20.2 & 1.4 & 0.2 & 2.9 & 0.3 & 2.7 & 0.9 & ~ &  & ~ &  \\ \hline 
37.4 & 0.6 & 102.9 & 11.6 & 177.6 & 20.5 & 182.9 & 20.5 & 1.1 & 0.2 & 3.8 & 0.4 & ~ &  & 1.9 & 0.3 & 9.9 & 3.0 \\ \hline 
35.7 & 0.6 & 111.4 & 12.6 & 179.5 & 20.7 & 177.2 & 19.9 & 1.3 & 0.2 & 3.2 & 0.4 & ~ &  & 1.7 & 0.3 & 12.6 & 3.8 \\ \hline 
34.0 & 0.6 & 124.9 & 14.1 & 201.5 & 23.0 & 169.9 & 19.1 & 1.2 & 0.2 & 4.3 & 0.5 & ~ &  & 2.5 & 0.3 & ~ &  \\ \hline 
32.2 & 0.7 & 123.1 & 13.9 & 198.4 & 22.8 & 161.5 & 18.1 & 1.2 & 0.2 & 4.2 & 0.5 & ~ &  & 1.9 & 0.3 & 23.2 & 3.8 \\ \hline 
30.3 & 0.7 & 112.9 & 12.7 & 166.1 & 18.9 & 164.4 & 18.5 & 1.0 & 0.2 & 3.4 & 0.4 & ~ &  & 2.3 & 0.3 & 10.9 & 6.4 \\ \hline 
28.4 & 0.8 & 146.6 & 16.5 & 184.0 & 20.9 & 165.9 & 18.6 & 1.4 & 0.2 & 3.8 & 0.4 & ~ &  & 2.7 & 0.4 & 11.3 & 2.4 \\ \hline 
26.3 & 0.8 & 148.7 & 16.7 & 188.9 & 21.7 & 157.7 & 17.7 & 1.0 & 0.2 & 3.4 & 0.4 & ~ &  & 3.0 & 0.4 & ~ &  \\ \hline 
24.1 & 0.9 & 135.5 & 15.3 & 163.5 & 18.7 & 131.7 & 14.8 & 1.4 & 0.2 & ~ &  & ~ &  & 2.1 & 0.3 & ~ &  \\ \hline 
21.7 & 0.9 & 171.0 & 19.2 & 156.9 & 17.8 & 137.8 & 15.5 & 0.9 & 0.2 & 1.6 & 0.2 & ~ &  & 3.40 & 0.41 & 29.4 & 5.5 \\ \hline 
19.2 & 1.0 & 166.3 & 18.7 & 152.3 & 17.5 & 101.2 & 11.4 & 0.44 & 0.13 & 1.6 & 0.2 & ~ &  & 2.69 & 0.34 & 15.8 & 2.7 \\ \hline 
16.3 & 1.0 & 146.2 & 16.5 & 145.6 & 16.7 & 35.1 & 3.9 & 0.25 & 0.07 & 2.2 & 0.2 & ~ &  & 2.99 & 0.35 & 13.8 & 1.9 \\ \hline 
14.2 & 1.1 & 164.4 & 18.5 & 173.4 & 19.8 & 14.9 & 1.7 & 0.10 & 0.02 & 2.2 & 0.3 & ~ &  & 2.77 & 0.35 & ~ &  \\ \hline 
11.8 & 1.1 & 69.0 & 7.8 & 89.0 & 10.3 & 0.25 & 0.04 & ~ &  & 1.8 & 0.2 & ~ &  & 1.47 & 0.19 & 6.1 & 1.0 \\ \hline 
10.3 & 1.2 & 46.1 & 5.2 & 61.8 & 7.2 & 0.23 & 0.04 & ~ &  & 1.3 & 0.1 & ~ &  & 1.03 & 0.14 & 4.4 & 0.7 \\ \hline 
9.5 & 1.2 & 32.8 & 3.7 & 41.1 & 4.8 & 0.19 & 0.03 & ~ &  & 0.77 & 0.09 & ~ &  & ~ &  & 3.0 & 0.9 \\ \hline 
8.6 & 1.2 & 20.2 & 2.3 & 22.0 & 2.8 & 0.26 & 0.04 & ~ &  & 0.46 & 0.05 & ~ &  & 0.40 & 0.07 & 2.9 & 0.5 \\ \hline 
7.6 & 1.2 & 9.2 & 1.0 & 11.4 & 1.4 & 0.19 & 0.03 & ~ &  & 0.31 & 0.04 & ~ &  & 0.15 & 0.03 & ~ &  \\ \hline 
6.6 & 1.2 & 3.8 & 0.4 & 4.1 & 0.5 & 0.26 & 0.03 & ~ &  & 0.17 & 0.02 & ~ &  & ~ &  & ~ &  \\ \hline 
5.4 & 1.3 & 1.3 & 0.2 & 1.7 & 0.2 & 0.19 & 0.02 &  &  &  &  &  &  &  &  &  &  \\ \hline 
\end{tabular}
\end{center}
\end{table*}

\begin{table*}[t]
\tiny
\caption{Measured cross-sections of the ${}^{144,143}$Pm radioisotopes}
\centering
\begin{center}
\begin{tabular}{|p{0.3in}|p{0.3in}|p{0.4in}|p{0.4in}|p{0.4in}|p{0.4in}|} \hline 
\multicolumn{2}{|p{1in}|}{\textbf{Energy\newline }(MeV) } & \multicolumn{2}{|p{0.8in}|}{\textbf{${}^{144}$Pm \newline }$\sigma\pm\Delta\sigma$(mb)} & \multicolumn{2}{|p{0.8in}|}{\textbf{${}^{143}$Pm \newline }$\sigma\pm\Delta\sigma$(mb)} \\ \hline 
49.2 & 0.3 & 4.9 & 0.7 & 38.6 & 4.4 \\ \hline 
47.9 & 0.3 & 6.3 & 0.8 & 37.9 & 4.3 \\ \hline 
46.5 & 0.4 & 6.9 & 0.9 & 37.7 & 4.4 \\ \hline 
45.1 & 0.4 & 5.6 & 0.7 & 38.4 & 4.4 \\ \hline 
43.6 & 0.4 & 6.6 & 0.8 & 40.6 & 4.6 \\ \hline 
42.1 & 0.5 & 5.5 & 0.7 & 41.6 & 4.7 \\ \hline 
40.6 & 0.5 & 6.7 & 0.9 & 38.7 & 4.5 \\ \hline 
39.0 & 0.5 & 6.5 & 0.8 & 41.8 & 4.8 \\ \hline 
37.4 & 0.6 & 5.8 & 0.8 & 47.4 & 5.4 \\ \hline 
35.7 & 0.6 & 4.6 & 0.7 & 47.1 & 5.4 \\ \hline 
34.0 & 0.6 & 5.5 & 0.8 & 44.6 & 5.1 \\ \hline 
32.2 & 0.7 & 4.2 & 0.6 & 38.0 & 4.3 \\ \hline 
30.3 & 0.7 & ~ &  & 33.3 & 3.9 \\ \hline 
28.4 & 0.8 & 4.5 & 0.6 & 39.3 & 4.5 \\ \hline 
26.3 & 0.8 & 2.7 & 0.4 & 24.0 & 2.7 \\ \hline 
24.1 & 0.9 & 3.8 & 0.6 & 6.0 & 0.7 \\ \hline 
21.7 & 0.9 & ~ &  & 7.8 & 1.2 \\ \hline 
19.2 & 1.0 & 1.1 & 0.2 & 4.6 & 0.6 \\ \hline 
16.3 & 1.0 & 0.7 & 0.1 & 1.6 & 0.3 \\ \hline 
14.2 & 1.1 & ~ &  & 1.6 & 0.7 \\ \hline 
11.8 & 1.1 & ~ &  & 0.20 & 0.27 \\ \hline 
10.3 & 1.2 & ~ &  & 0.44 & 0.11 \\ \hline 
9.5 & 1.2 & ~ &  & ~ &  \\ \hline 
8.6 & 1.2 & ~ &  & ~ &  \\ \hline 
7.6 & 1.2 & ~ &  & ~ &  \\ \hline 
6.6 & 1.2 & ~ &  & 0.24 & 0.06 \\ \hline 
5.4 & 1.3 & ~ &  & 0.32 & 0.05 \\ \hline 
\end{tabular}
\end{center}
\end{table*}

\subsection{Integral yields}
\label{4.2}
The integral yields as a function of the incident deuteron energy was calculated from a cubic spline fit to our experimental data and are shown in Fig. 23 and Fig. 24 for Eu and Sm, Pm radionuclides respectively. The values represent so called physical yield, considering an instantaneous short irradiation and measuring immediately after EOB allowing to neglect all corrections for decay \cite{12}. The calculated integral yields for $^{148}$Eu,$^{150m}$Eu and $^{154g}$Eu are compared with the only experimental results of Dmitriev et al. \cite{3}. In the case of $^{148,150,151}$Eu the agreement is acceptable.

\begin{figure}
\includegraphics[scale=0.31]{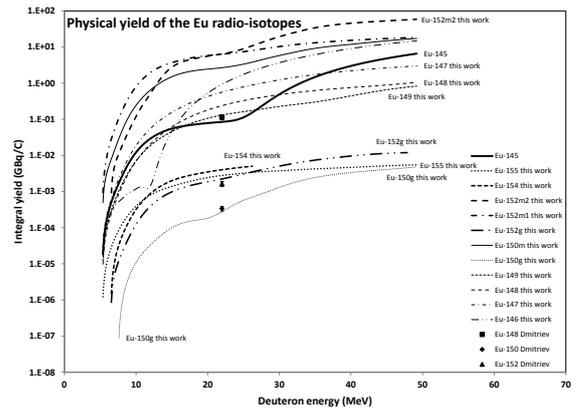}
\caption{Thick target yields for radionuclides of europium produced by deuteron irradiation on $^{nat}$Sm }
\end{figure}

\begin{figure}
\includegraphics[scale=0.31]{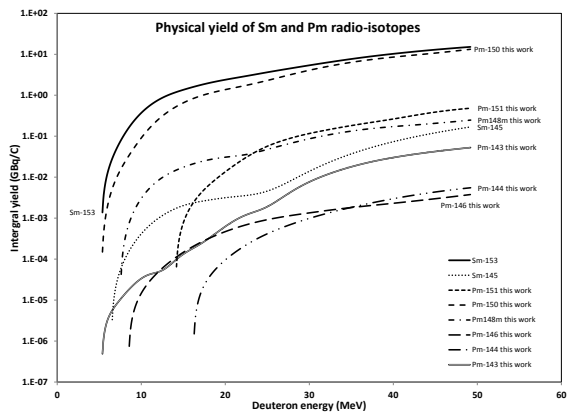}
\caption{Thick target yields for radionuclides of samarium and promethium a produced by deuteron irradiation on $^{nat}$Sm}
\end{figure}

\section{Summary and conclusion}
\label{5}
We present experimental  cross-sections $^{nat}$Sm(d,xn)$^{155,154,152m2,152m1,152g,150m,150g,149,148,147,146}$Eu, $^{nat}$Sm(d,x)$^{153,145}$Sm and $^{nat}$Sm(d,x)$^{151,150,149,148m,148g,145,144,143}$Pm up to 50 MeV measured for the first time. Our excitation functions are relative to recommended monitor reactions excitation functions, re-measured simultaneously over the whole covered energy range, and are hence very accurate concerning energy scale and number of bombarding particles. 
The experimental data were compared with the results of our theoretical calculations using the ALICE-D and EMPIRE-D codes and with results in the TENDL-2012 on-line library based on the latest version of the TALYS code. The predictions of the model codes are quite different and disparate. Our experimental data are in average lower than the results of the theoretical codes, with the largest overestimation for EMPIRE-D. The systematic disagreement with the experimental data underlines the importance of the experimental data as basis for improvement of description of reaction mechanisms in the codes and for upgrading of the input parameter libraries. 
The measured data can be useful in for determining activation in different applications of samarium where charged particle irradiation fields are present:
\begin{itemize}
\item	It is used in magnets, in alloys with cobalt
\item	It is used in airspace and defense industries (samarium permanent magnets that are stable at high temperatures, precision-guided weapons, "white noise" production in stealth technology.
\end{itemize}
Two isotopes of Sm are relevant in nuclear medicine: $^{145}$Sm is used for brachytherapy and $^{153}$Sm for treatment of bone pain. Although these radioisotopes are mostly produced in research reactors by neutron caption on samarium (enriched targets if high yields and radionuclidic purity is needed) low energy deuteron induced reactions on enriched $^{144}$Sm or $^{152}$Sm can be an alternative for local production. 
$^{149}$Sm is a strong neutron absorber and it is therefore added to the control rods of nuclear reactors. Knowledge of the cross-sections of the activation products induced by secondary products of neutron reactions has also importance.
 
\section{Acknowledgements}
\label{6}

This work was done in the frame MTA-FWO research project. The authors acknowledge the support of research projects and of their respective institutions in providing the materials and the facilities for this work.
 



\bibliographystyle{elsarticle-num}
\bibliography{Smd}







\end{document}